\documentclass[aps,showpacs,prb,twocolumn,supperscriptaddress]{revtex4-1}
\usepackage{graphicx} 
\usepackage{dcolumn} 
\usepackage{bm} 
\usepackage{graphicx,color,epsfig,rotate}
\usepackage{amsmath,amssymb}

\begin{document} 

\title{ Partial Equilibration Scenario in 3D athermal martensites\\ quenched  below first-order transition temperatures}
\author{N. Shankaraiah${^1}$ , K.P.N. Murthy${^2}$ and S.R. Shenoy${^1}$ }
\affiliation{ ${^1}$Tata Institute of Fundamental Research-Hyderabad, Hyderabad, Telangana 500046, India. \\
${^2}$Dept of Physics, Central University of Rajasthan, Bandar Sindri, Rajasthan 305817, India}

\date{\today}

\begin{abstract}

 \noindent   To test a  Partial Equilibration Scenario (PES) of Ritort and colleagues, we do Monte Carlo simulations of discretized-strain spin models, for four 3D martensitic structural transitions under quenches to a bath temperature $T <T_0$  below a first-order transition. The ageing system faces entropy barriers, in {\it searches} for energy-lowering passages between quasi-microcanonical energy shells. We confirm the PES signature of  an exponential-tail distribution of intermittent heat releases to the bath, scaled in an effective temperature, that  in our case, depends on the quench.  When its inverse $\beta_{eff} (T) \equiv 1/T_{eff} (T) $ vanishes below a `martensite start' temperature  $T_1$ of avalanche conversions, then entropy barriers vanish.  When this search temperature $T_{eff} (T)$ vanishes, PES cooling  is arrested, as entropy barriers diverge. We find a {\it  linear} vanishing of   $T_{eff}(T)\sim T_d -T$, below a  delay-divergence temperature $T_d$ in between, $T_1 < T_d < T_0$.   Martensitic conversion delays  $e^{1/T_{eff}} \sim e^{1/(T_d -T)}$  thus  have  Vogel-Fulcher-Tammann like divergences. Post-quench delay data  extracted  from simulations and  athermal martensitic alloys, are both consistent with predictions.   

\end{abstract}
\maketitle

\vskip 0.5truecm

\section{Introduction}

The  re-equilibration of a system after a quench is a long-standing  problem in non-equilibrium statistical mechanics,  and a generic Partial Equilibration Scenario (PES)  has been proposed  by Ritort and colleagues \cite{R1,R2,R3,R4} After a quench, the system  spreads rapidly over an energy shell $E$  in configuration space. The system lowers its energy by intermittent energy releases $\delta E <0$ to the heat bath at $T$, and then  again spreads ergodically over the next energy shell  $E' = E + \delta E < E$. The iteration of the  fast/ slow steps ratchets  the system down to a canonical equilibrium at $T$.The distribution of energy changes $\{\delta E\}$ has a signature exponential tail \cite{R1,R2,R3,R4} in the heat release  distribution 
 $P_0(\delta E; t_w)\sim e^{\delta E / 2 T_{eff} (t_w)}$, whose slope  at the origin defines the  (inverse) effective temperature $\beta_{eff} \equiv 1/T_{eff}$, dependent on the post-quench waiting time $t_w$. The tail for negative energies is part of a  shifted gaussian that peaks at positive energies.  The PES has been confirmed through analytic Monte Carlo (MC) calculations of   relaxing  independent harmonic oscillators \cite{R1,R2}; by simulations of spin-glass models and Lennard-Jones binary mixtures\cite{R3,R4}, and through voltage noise intermittency  \cite{R3}. The MC updates of ageing  harmonic oscillators \cite{R2}   have to hit the ever-shrinking target of  as yet unrelaxed oscillators, and these rising entropy barriers induce slow decreases in  energies $1/\ln t$ and acceptances $\sim 1/t\ln t$. 

 It is natural to test PES ideas in interacting systems with slow relaxations. The structural glass transition  \cite{R5} is an attempted equilibration  that is  arrested at a glassy freezing temperature  $T_G$, pre-empting crystallization \cite{R5,R6}.  The $T$-dependent effective viscosity increases above a Vogel-Fulcher-Tammann (VFT) singularity \cite{R7,R8} $ e^{1/|T-T_G|}$, that has been studied for a century, but is not yet fully understood. Further, decays around the glass temperature have non-exponential time dependence; and non-Debye frequency responses \cite{R9}.
 We  consider other  systems with structural transitions and equilibration delays.  Martensitic steel alloys,  when quenched from high temperature parent austenite to low temperature `martensite'  \cite{R10,R11},  show  strain  domain-wall\cite{R12} patterns \cite{R13,R14,R15}.They  can exhibit puzzling delays in conversion to  martensite  \cite{R16,R17,R18,R19,R20}, that  increase rapidly with temperature: raising $T$ to nearer transition by  a few percent, can raise delays from $1$ sec  to $10^4$ seconds \cite{R18}.

We do Monte Carlo  (MC) simulations of  quenches in $T$, of martensitic discretized-strain\cite{R21,R22,R23} models in {\it 3D}. The model hamiltonians  describe the elastic Domain Walls (DW) or mobile twin boundaries,  of  four  3D structural transitions, each with characteristic anisotropic Compatibility interactions between order-parameter strains\cite{R15}. The four transitions \cite{R21} can occur in martensite-related functional materials, where OP strains will be coupled to the functionality variables. The transitions are tetragonal-orthorhombic (YBCO, superconductivity); cubic-tetragonal (FePd, shape memory);  cubic-orthorhombic (BaTiO, ferroelectrics); cubic trigonal (LaSrMnO, colossal magnetoresistance). 

  Our 3D simulations yield  post-quench evolutions as in \cite{R24,R25,R26}  2D,  passing through three Domain Wall (DW) states. At first there is a majority-austenite  DW Vapour  state of  a martensite droplet in an austenite background.  This converts to majority-martensite  DW Liquid,  of  randomly wandering walls. Finally the DW Liquid  orders to a DW Crystal microstructure, with the  walls  along preferred directions \cite{R13},  as in  parallel  `twins'. 
  
  We  focus on the conversion delays of the first evolution of  DW Vapour $\rightarrow$ DW Liquid, that corresponds to austenite to martensite conversions, or a rise of the martensite fraction $n_m (t)$ from zero  to unity. As in the earlier  2D  case  \cite{R24,R25,R26},  a phase diagram in material parameters is obtained. In the  `athermal' martensite regime, there are curious  `incubations', or   no apparent changes after a quench, terminated by  sudden avalanche conversions  \cite{R16,R17,R18,R19,R20}. In this regime,  we find  three characteristic temperatures, with $T_1 < T_d < T_0$. Here  $T_0$ is the meanfield transition to uniform ordering.

 Avalanche conversions  in  a single MC time step $t_m  =1$,occur  for $T< T_1$, identified as the martensite start temperature\cite{R10,R11} $ M_s= T_1$. 
  Quenches into $T_1 < T < T_d$ show (postponed) avalanches \cite{R27} or {\it  incubation }behaviour: the conversion fraction $n_m(t)$ remains flat at zero, until there is a jump up to unity a  time $t=t_m(T)$. The incubation delay time $t_m (T)$ extends rapidly, on approaching a divergence temperature $T_d$.The physical picture for delays is of Vapour-droplet Fourier profile attempting entry to a negative-energy region of effective Hamiltonian spectra $\epsilon (\vec k, T) <0$.  The profile has to pass through a zero-energy $\vec k$-space contour at a  bottleneck, like a  Golf Hole edge \cite{R28,R29,R30,R31}. This transit passage delay differs from the familiar critical slowing down from a divergent Order Parameter  length. The  $T$-dependent, anisotropic  bottleneck shrinks on warming, with a topological shape change at $T_d$, that blocks entry, so entropy barriers diverge. The precursor\cite{R14,R15}  region $T_d < T < T_0$, where 2D results suggest dynamic tweed \cite{R27},  may be studied elsewhere.

 We use generic equilibration scenario  of Ritort and colleagues to analyze the statistics \cite{R1,R2,R3,R4,R31}, of the set of energy changes $\{\delta E\}$ from each MC step  (usually used but not retained \cite{R32}).
These heat releases are recorded  only up to an aging time $t_w =t_m (T)$, so the effective temperature  depends on the quench temperature, $T_{eff}(t_w) \rightarrow  T_{eff} (T)$: non-stationary distributions become time independent.  We confirm the signature PES exponential tail for all four transitions. We find the search temperature  vanishes linearly  $T_{eff}(T) \sim T_d-T$, driving an entropy barrier divergence at $T_d$. The martensite-conversion times  are thus predicted to show  glass-like VFT behaviour,  $\ e^{1/T_{eff} (T)}\sim e^{1/(T_d -T)}$ here understood as an arrest of PES cooling. Such VFT behaviour,  is extracted from simulation and experimental \cite{R16,R17,R18} data. 
 \begin{figure}[h]
\begin{center}
\includegraphics[height=4.5cm,width=8.0cm]{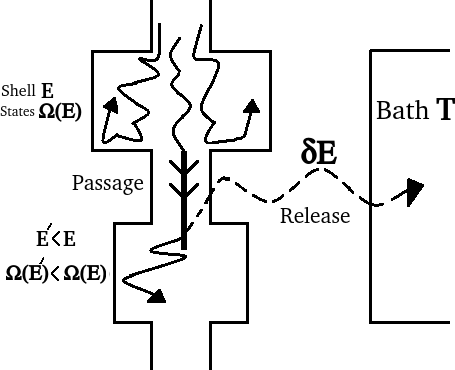}
\caption{{ \it  Schematic of Partial Equilibration Scenario :} The system  in a configurational  shell of states  with energy $E$, states $\Omega(E)$, and entropy $S(E) = \ln \Omega (E)$ makes a passage to the next shell of lower $E' < E$,   fewer states  $\Omega (E')< \Omega (E)$ and smaller entropy $S(E') < S(E)$, crossing  a generic entropy barrier to find rarer states. The turned-back  wiggly lines are failed searches for passage. The downward  bold line denotes successful searches, accompanied by a distribution of  heat releases $\delta E \equiv E'-E < 0$  to the bath at $T$, scaled in an effective search temperature $T_{eff}$. Our case of quenches below a first-order transition can have  further, specific entropy barriers, eg  hindering system  passage through   order parameter-related, $T$-dependent, bottlenecks in  phase-space.  }
\end{center}
\end{figure}

The plan of the paper is as follows. In Section II, we discuss the generic Partial Equilibration Scenario and our  specific case of quenching across a phase transition. Section III describes for the four transitions in 3D, the  discrete-strain clock-like spins and their  $T$-dependent Hamiltonians.  In Section IV we describe the MC simulations, with delay results from the phase space bottlenecks in Section V.  Section VI shows that PES signatures are seen in all four transitions. Section VII shows that  both 3D simulations and metallic alloy experiments exhibit VFT  behaviour. Finally Section VIII is a  summary. 

Appendix A illustrates how a continuum double-well Landau free energy induces a T-dependent,  Ising effective Hamiltonian.  Appendix B obtains the  athermal phase diagram for  four transitions. Movies of post-quench DW evolutions are in Supplementary Material Videos \cite{R27}.

\begin{figure*}
\begin{center}
\includegraphics[height=4.0cm, width=16.0cm]{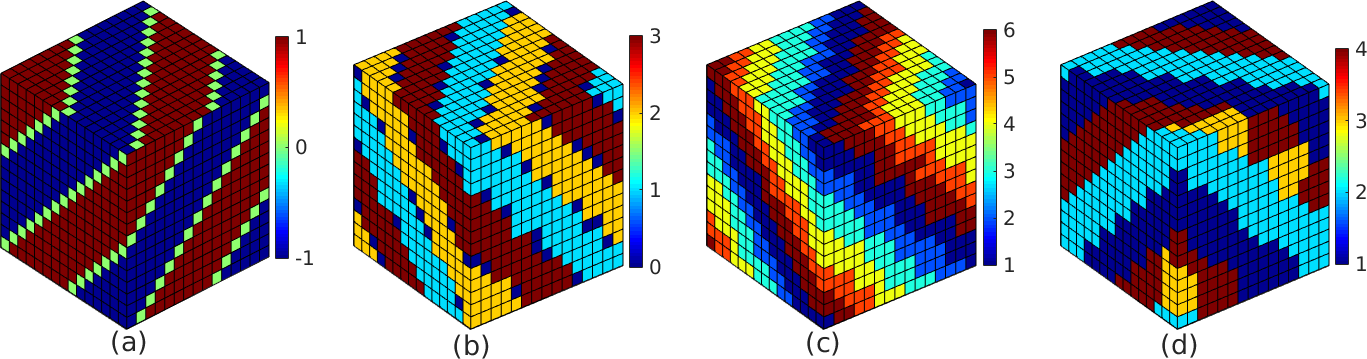}
\caption{{\it Microstructures in ferroelastic transition:} The Domain Wall crystal or twinned textures are shown, with variant label $V$ in the colour bar. The Hamiltonian energy scale in is $E_0 =3$,  with the non-OP compressional constant $A_1=4$ fixing all other  non-OP elastic constants, see text. The DW textures are (a) tetragonal-orthorhombic transition at $T=0.45$ and Landau spinodal temperature $T_c=0.95$;  (b)  cubic-tetragonal transition at $T=0.4$ and $T_c=0.95$;  (c) cubic-orthorhombic transition at $T=0.3$ and  $T_c=0.95$;  and (d) cubic-trigonal transition at $T=0.37$ and $T_c=0.97$. Note that  for each transition, all allowed variants are actually present. }
\end{center}
\end{figure*}

\section{ Scenario for post-quench equilibration:}

How do systems re-equilibrate, after a temperature quench?
 Ritort and coworkers have suggested that if an equilibrium canonical ensemble in thermal contact with a heat bath suddenly has its bath  quenched to a lower temperature $T$, then the system goes into an ageing ensemble\cite{R1,R2,R3,R4}, that has a quasi- microcanonical description of states of the system. There are  sequential passages through  decreasing-energy configurational shells, and intervening  entropy barriers in the system-search for the new equilibrium. While delays from energy barriers  are from attempts through activated jumps, to cross mountains, delays from entropy barriers are from attempts through constant-energy searches, to  find rare  channels going  through  or around, the mountain \cite{R31}.

 In this Section we i)  outline (our understanding  of) the generic  Partial Equilibration Ensemble \cite{R1,R2,R3,R4} or PES, and ii) state how this  ageing scenario is applied to our specific case, that has  quenches across phase transitions, and  order parameter emergence from zero.\\

{\it i) Generic  PES for ageing after quenches: }
 
The PES for the equilibration process considers  a system of energy $E$ in contact with a  (larger) heat bath. A  familiar textbook derivation \cite{R33} of the canonical ensemble, applies the microcanonical ensemble to the system plus heat bath,  of constant total  energy $E_{tot} = E +E_{bath}$. The total number of states is the product of the $\Omega(E)$ states of the system, and of the  $\Omega_{bath}(E_{tot} -E)$ states of the bath, summed over all  allowed system energies $0< E <E_{tot}$:
\begin{equation} 
\Omega_{tot} (E_{tot}) = \sum _{E} \Omega (E) \Omega_{bath} (E_{tot} - E) .
\end{equation}

 The system configurational  entropy  is $S(E) = \ln \Omega(E)$,  and  the inverse effective  temperature is  $\beta_{eff} \equiv 1/ T_{eff} = dS(E)/ dE$.  Similarly, the inverse bath temperature is $\beta_{bath} \equiv 1/T_{bath} = dS_{bath}/dE_{bath}$. The change in  total  entropy depends on the system energy change $d E$ as
  
\begin{equation}
 dS_{tot} = \frac{dS (E)}{dE} dE +\frac{dS_{bath} }{dE_{bath}} dE_{bath} = [\beta_{eff}  - \beta_{bath}] dE \geq 0
\end{equation}
with the equality at equilibrium, when the system and bath temperatures are equal $ \beta_{eff}  = \beta_{bath}$. The Second Law  inequality  $dS_{tot} >0$  must hold, for the  irreversible cases.  After a bath quench from initial to a final $T_{bath} =T$,  the system is  left hotter,  $T_{eff} > T$, or $\beta_{eff} -\beta < 0$.  The  energy changes  are negative $dE= dQ  <0$, with heat released {\it  by}  the system {\it to} the bath. 

At equilibrium, the  terms in the sum of Equ (1) are dominated by clusters of energy shells selected \cite{R33} by a sharp peak, arising from the product of  a rising number of system states $\Omega (E)$, and a falling  bath factor $e^{-E/T_{bath}}$. The  peak width is the energy fluctuations from stochastic system-bath  exchanges.  These equilibrium ideas describe {\it states},  not processes. 

The Partial Equilibration Scenario postulates a plausible  post-quench  non-equilibrium {\it  process} for the system to evolve between the initial and the final equilibrium state. A sudden change in  bath temperature or quench, will induce a {\it shifted} peak,  around a different equilibrium state.  The post-quench system, initially stranded in non-optimum states, is visualized as moving through the sequentially lower energy shells of Equ (1) in its search for the shifted peak,  tracked by an ageing  time $t_w$.

 The Scenario postulates that the system  i) rapidly  spreads ergodically through  all states of a  shell of energy E, and ii) slowly dribbles out  energy  $\{\delta E\}$  to the ever-present energy bath. Since the system is partially equilibrated in the quasi-microcanonical  shell, the equilibrium definitions can be retained, of the shell entropy $S(E)$ and its energy derivative $1/ T_{eff}(t_w)$. The back-and-forth energy  exchanges to rapidly surmount internal energy barriers and  explore all shell configurations, are summoned by the system from the bath (`stimulated'). The slow changes on  passages to a lower-energy shell, are releases by the system to the bath ('spontaneous'). 
 
  Fig 1 is a schematic of  the Partial Equilibration Scenario. The successive shells have lower energy $E' < E$ and hence lower number of configurations $\Omega (E') < \Omega (E)$ and entropies $S(E') < S(E)$. There is a generic  {\it entropy} barrier  $S_B \equiv -\Delta S = -\ln [ \Omega (E') / \Omega (E)] > 0$ to finding the rarer states. Key seeks lock: most attempts fail.  
   
From an ageing Fluctuation Relation \cite{R3} the nonequilibrium  energy-change probability \cite{R1,R2,R3,R4} is a peak at the origin,  times an exponential tail for negative changes. This generic PES signature tail depends on the ageing time through the
 the effective temperature $T_{eff} (t_w)$, that scales the heat releases:
\begin{equation}
P_0(\delta E; t_w) \simeq  {P^{(+)}}_0 (\delta E; t_w) e^{ \delta E/2 T_{eff} (t_w)}, 
\end{equation}
with an even  prefactor ${P^{(+)}}_0 (\delta E;t_w)$. 

In an important result, other effective temperatures, from the Fluctuation-Dissipation Theorem;  and from non-equilibrium  fluctuations of system  variables,   are  shown to be equal to  the PES effective temperature \cite{R1}:  there is only one $T_{eff}$. 
   
 \begin{figure}[ht]
\begin{center}
\includegraphics[height=6.5cm, width=8.5cm]{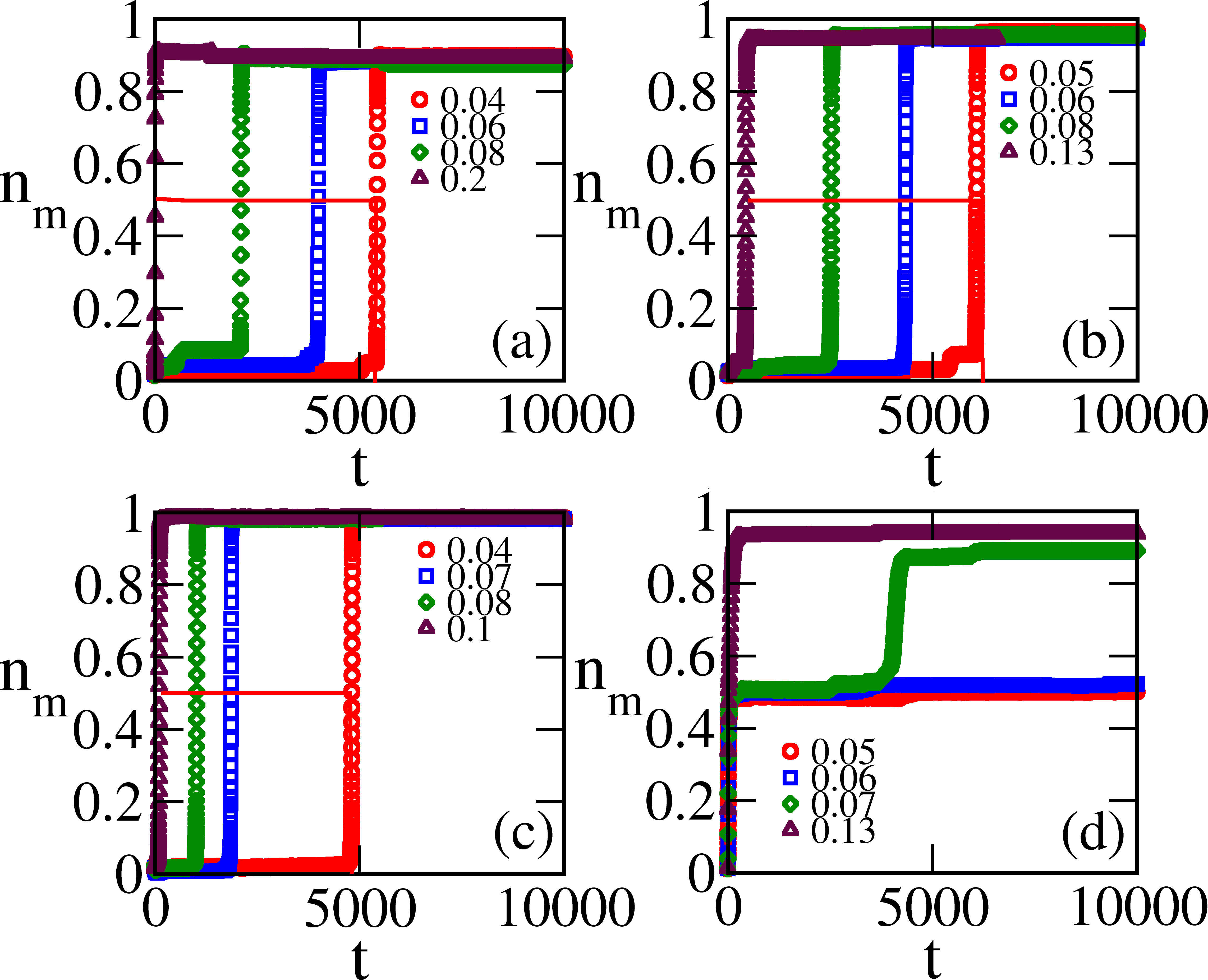}
\caption{{\it Conversion incubations ending in postponed avalanches:} Four-panel plot of martensite conversion fraction $n_m(t)$ versus time $t$, for  fixed  $E_0=3$, $A_1=4$ and different scaled temperatures $|\delta_0 (T )| \equiv |T-T_d|/ T_d$ as in  the legend. There are   flat incubations ending in explosive jumps in $n_m(t)$ at $t= t_m$ defined by $n_m (t_m)=1/2$. (a) tetragonal-orthorhombic transition with Landau spinodal temperature $T_c=0.9$; (b) cubic-tetragonal transition with $T_c=0.95$; (c) cubic-orthorhombic transition with $T_c=0.95$;  (d)  cubic-trigonal transition with $T_c=0.97$, that is unusual, see text. }
\end{center}
\end{figure}

 \begin{figure}[ht]
\begin{center}
\includegraphics[height=6.5cm, width=8.0cm]{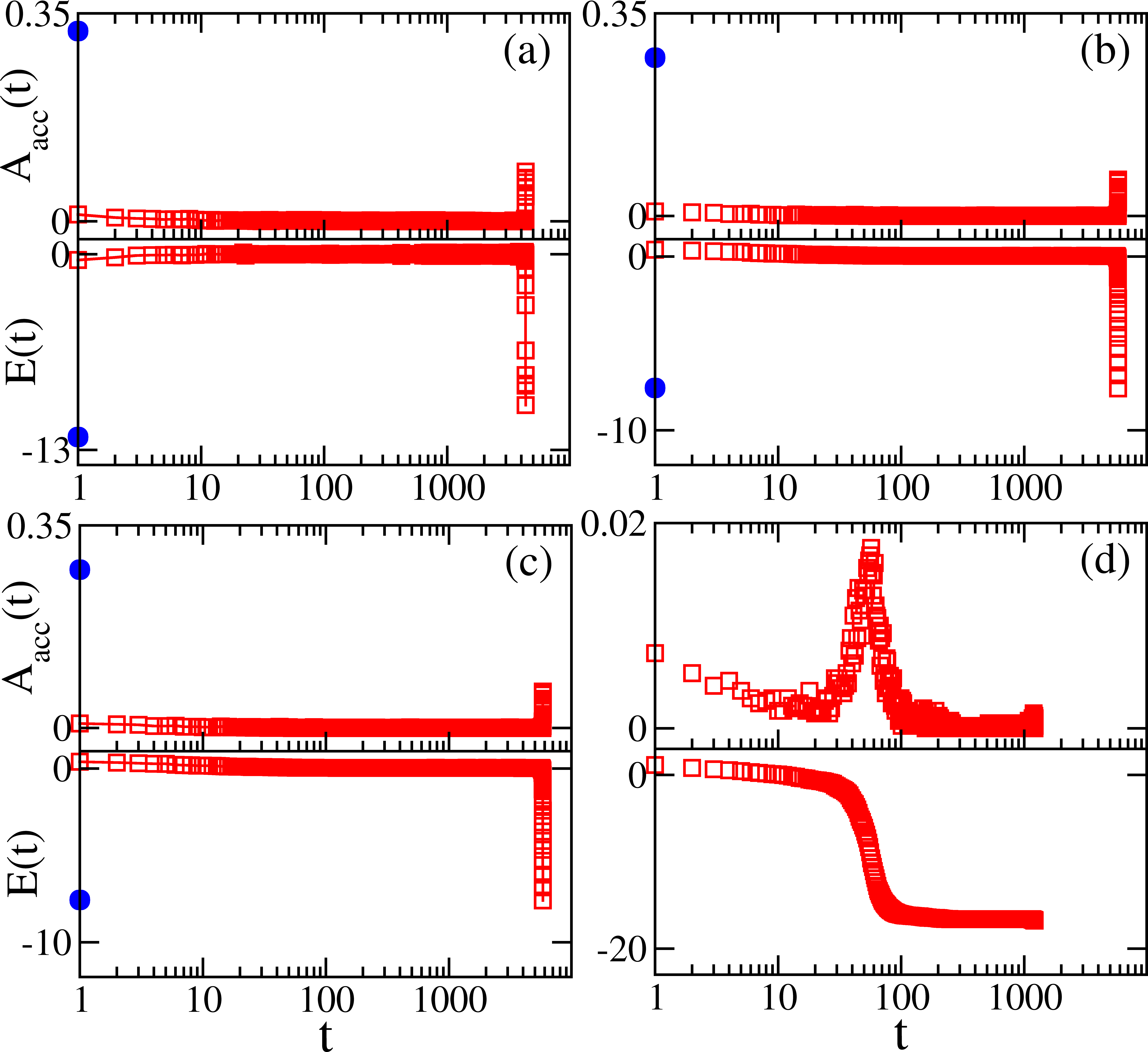}
\caption{{\it Acceptance spikes and energy drops:} Four-panel plots for all transitions,  showing  Acceptance fractions $A_{acc}(t)$ and total energies $E(t)$ versus time (displaced above and below the $x$-axis, for clarity).  For  $T <T_1$, eg  for  $\delta_0  =-0.7$,  there are immediate spikes in $A_{acc} (t)$, and drops  to negative values in energy  $E (t)$,  at the  very first MC sweep $t=1$ as denoted by  on-axis blue solid circles. For $ T > T_1$  eg for $\delta_0 =-0.1$, the red open squares show  there are  flat lines of incubation ending at  $t=t_m$ postponed spikes and drops. The transitions a), b), c), d) are as in the Fig 3 Caption. Again,  the cubic-trigonal case d) is unusual. }    
\end{center}
\end{figure}

{\it  ii) Specific  PES from quenching across a transition.}

For our case of quenching to $T$ across a first-order transition, the Order Parameter (OP) has to  rise from zero, and  so  the  wait times $t_w$ to reach  OP marker events will  depend on the quench temperature, $t_w=t_m(T)$.The effective temperature   and PES distribution will thus also depend on the quench temperature,  $T_{eff} (t_w) \rightarrow T_{eff} (T)$ and  $P_0 (\delta E; t_w)  \rightarrow P_0(\delta E, T)$.   The even prefactor $P^{(+)}_0 (\delta E,T)$ can be exponentiated and expanded  to  quadratic order,  Equ (3) is then a gaussian peaked at the origin, times an exponential falloff. Completing the square yields  a PES signature of a shifted gaussian, peaked at positive mean changes $M(T) = < \delta E>$, and scaled in $T_{eff}$:

\begin{equation}
P_0(\delta E, T) \simeq e^{-[\delta E - M(T)]^2 / 4 M(T) T_{eff} (T)}.
\end{equation}
 
 For small heat releases $\delta E = - |\delta E| <0$, the PES distribution  takes a  Boltzmann-like form $P_0 \simeq   e^{-\frac{1}{2} \beta_{eff}(T) |\delta E|}$. This gives a physical meaning to the effective temperature: it is a {\it search  range} for accessible energy shells. If  $\beta_{eff} \rightarrow 0$, entropy barriers collapse, and  passages are immediate.  If $T_{eff} \rightarrow 0$, then entropy barriers diverge, and  passage-searches freeze.

We postulate that  the OP-related  bottlenecks  can be of two  types, depending on the depth of the quench. a) The DW Vapour $\rightarrow$ Liquid delays are attributed to phase space bottlenecks \cite{R27}, suggested by concepts in protein folding \cite{R28,R29,R30}. Hamiltonian energy-spectrum contours in Fourier space  of zero energy are like a $T$-dependent Golf Hole (GH) edge, with a  negative-energy Funnel region inside it leading to the final state \cite{R28}.  The entropic delays  are from finding and entering  the bottleneck.   \\
 b) For deeper quenches, the DW Liquid $\rightarrow$  Crystal delays could be conceptually related to  spin facilitation models \cite{R27,R34,R35}, through the $T$-dependent sparseness of austenitic-hotspot dynamical catalysts,  or other facilitating  fields \cite{R20,R27}.

\section{ Domain-wall Hamiltonians for four structural transitions in 3D:}

The transition-specific, derived effective Hamiltonians have been  presented in detail\cite{R15,R21,R22,R23,R24,R25,R26}, and we just outline as conceptual background:
A) Strains and Compatibility constraints. B) Reduction of continuum strains to discrete-strain `pseudo-spins'. C) Reduction of  continuum strain free energies to effective `spin'  Hamiltonians.  

It is useful to define $N_{OP}$ the number of components of the OP strains; and $N_V$ the number of Landau `variant' minima at nonzero  OP strain  values. In terms of discretized strains,  $N_{OP}$ is the number of vector spin components, that can point in   $N_V$  variant directions. A double-well Landau free energy  for a scalar magnetization maps onto (Appendix A) an Ising model with $N_{OP} =1$ and  $N_V =2$.  We consider four  first-order transitions \cite{R21}  with $N_{OP} =1,2,2,3$. The nonzero, unit-magnitude variant vectors  point respectively  to corners of symmetry-dictated polyhedra with  $N_V=2,3,6,4$  corners,  inscribed in a unit circle or unit sphere: a geometrically  pleasing sequence of  line, triangle, hexagon,  and tetrahedron. These transitions are respectively, tetragonal-orthorhombic, cubic-tetragonal, cubic-orthorhombic, and cubic-trigonal.

\subsection{ Strains and Compatibility:}  

Strains  are symmetric tensors   ${\bf e} = {\bf e}^T $, where the superscript $T$ is Transpose. In three spatial dimensions,  there are  6 independent Cartesian strains \cite{R21,R23}  $e_{xx}, e_{yy}, e_{zz}, e_{xy}, e_{yz}, e_{zx}$. The  {\it physical} strains  $e_1,e_2...e_6$ are convenient linear combinations:   one compressional $e_1= (e_{xx} + e_{yy} + e_{zz})/\sqrt{6}$;  two deviatoric or rectangular $e_2 = (e_{xx} -e_{yy})/\sqrt{2}, ~e_3 =(2e_{zz} - e_{xx} -e_{yy})/\sqrt{6}$,  and three shears  $e_4 =e_{yz}, e_5=e_{zx}, e_6 = e_{xy}$. 

The free energy \cite{R21}   has a nonlinear Landau term that depends on a subset $N_{OP}$ of these physical strains, as the Order Parameter(s). The remaining $n = 6 - N_{OP}$  {\it non}-Order Parameter physical strains enter the free energy as harmonic springs, whose extensions  cannot be  simply be set equal to zero, as pointed out by Kartha \cite{R15} . This is because a local OP-strained unit cell will generate non-OP strains in surrounding unit cells. To maintain lattice integrity all strained  unit  cells must  mutually adapt,  to all fit together in a {\it smoothly compatible} way, without dislocations. 

For electromagnetism,  there is a  no-monopole Maxwell condition of vanishing divergence of the magnetic induction vector, $\nabla.{\vec B} =0$.  For  elasticity,  there is a no-dislocation  St Venant Compatibility condition  of a vanishing   double curl \cite{R15}, of the Cartesian strain tensor. In coordinate and Fourier space, 

\begin {equation}
\nabla \times [ \nabla \times {\bf e} (\vec r)]^{T} =\vec 0;\\
~~{\vec K} (\vec k) \times {\bf e} (\vec k)\times {\vec K} (\vec k) = \vec 0.
\end{equation}
Here  $K_\mu (\vec k) \equiv 2 \sin (k_\mu a_0 /2)$ for $\mu =x,y,z$, and lattice constant  $a_0 =1$.
There are  six differential-equation constraints, that are algebraic equations in Fourier space, of which only three are independent \cite{R21,R23}.
Going to  physical strains $e_1, e_2...e_6$  the three  $\vec k \neq 0$  algebraic equations express  the non-OP strains in terms of the OP strains.
The {\it uniform} $\vec k =0$ non-OP strains are not so constrained, and can be freely set to their minimum value of zero.

The harmonic  non-OP terms  can then be analytically  minimized subject to the $\vec k \neq 0$ linear constraints,  by direct substitution for non-OP strains  or by Lagrange multipliers. This yields an  OP-OP effective interaction, with a transition-specific  Compatibility Fourier kernel \cite{R15} that  depends on direction  $\hat k \equiv \vec k / |\vec k|$. The kernels all have a prefactor $\nu_{\vec k} \equiv 1 - \delta_{\vec k, \vec 0}$, that vanishes for  $\vec k =\vec 0$.  The Compatibility kernels are smallest (eg zero)  for specific directions $\hat k$,  explaining  the observed DW orientation along preferred crystallographic directions. The Compatibility potential in coordinate space is an anisotropic  powerlaw,  with  the  spatial dimensionality $d=3$ as the fall-off exponent $\sim 1/ R^d$.

\begin{figure}[ht]
\begin{center}
\includegraphics[height=6.5cm, width=8.5cm]{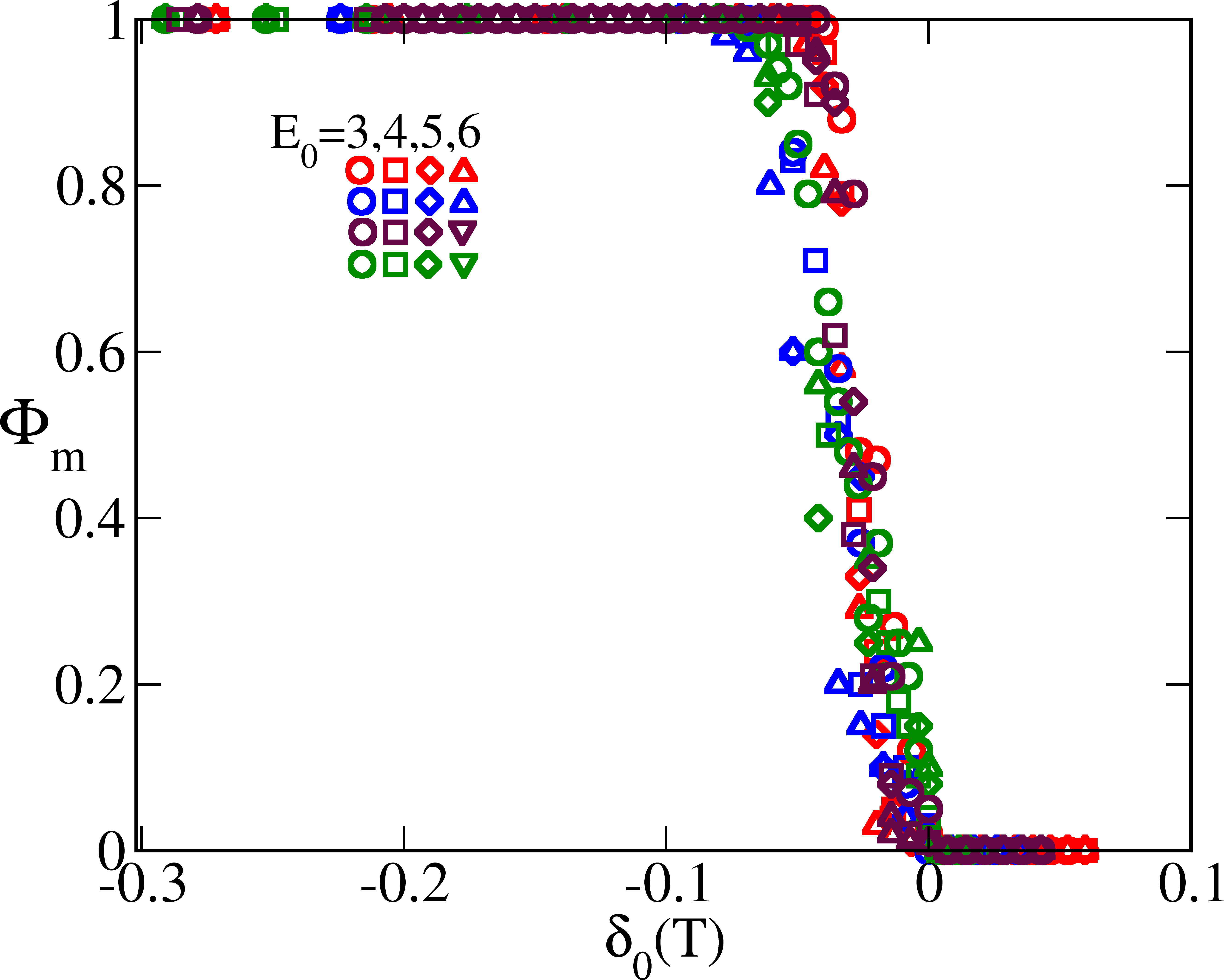}
\caption{{\it Conversion-success fraction:} The successfully converting  fraction  $ \Phi_m (T)$ over 100 runs versus $\delta_0 (T) \equiv (T- T_d)/ T_d$ is shown  in the range $T_1 <T < T_0$. The colours of symbols top to bottom denote transitions  in  order  a) tetragonal-orthorhombic, b) cubic-tetragonal, c) cubic-orthorhombic, and d) cubic-trigonal. For  a `precursor' region $T_0 > T> T_d$, conversions do not occur. Success fractions are not exponentially sensitive to Hamiltonian energy scales $E_0=3,4,5,6$, and are hence attributed  to {\it entropy} barriers.}
\end{center} 
\end{figure}

\subsection{ Discrete-strain pseudo-spins: }

\begin{figure}[ht]
\begin{center}
\includegraphics[height=6.5cm, width=8.5cm]{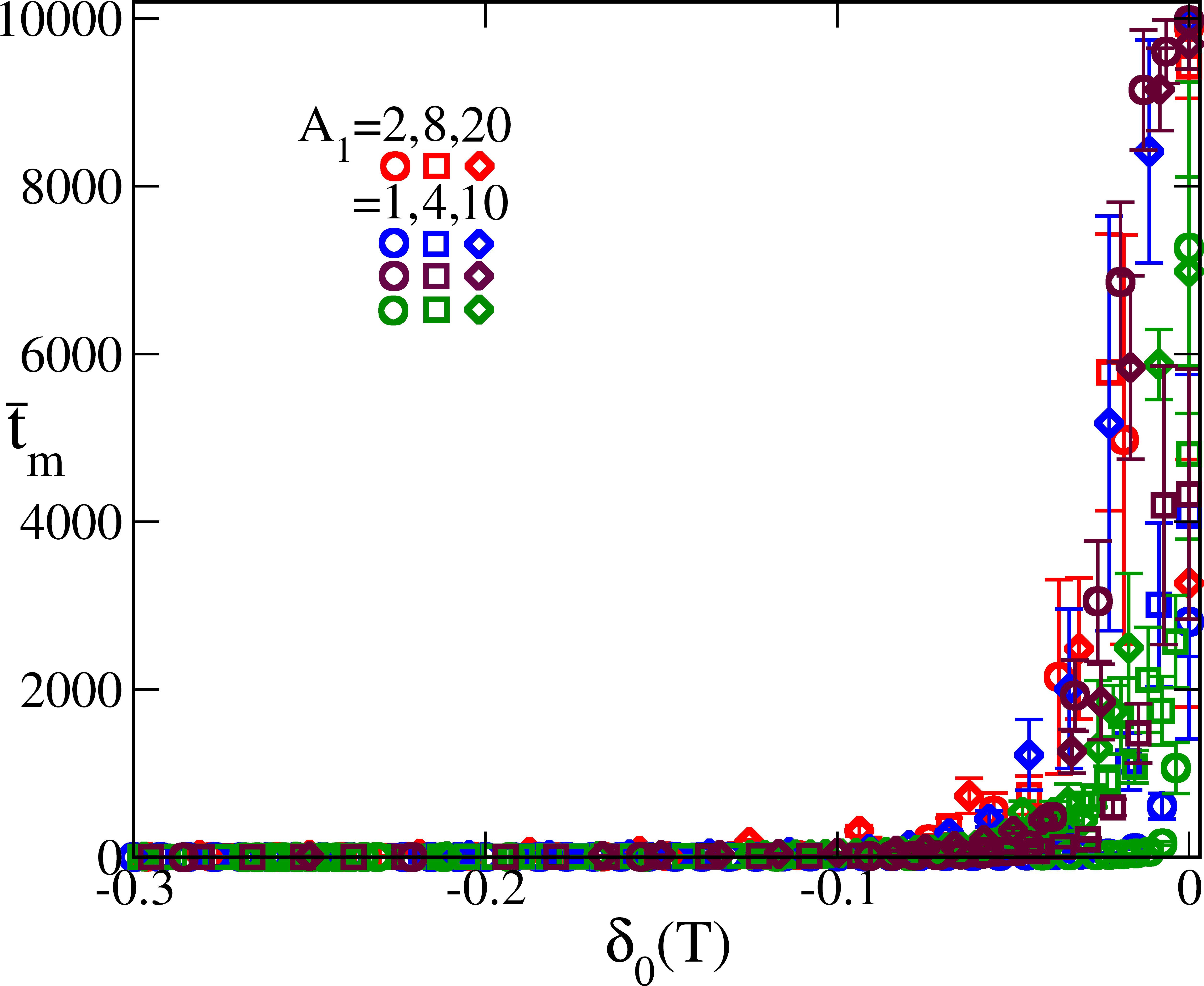}
\caption{{\it Delay times versus quench temperature:}  Linear-linear plots of mean  conversion delay time ${\bar t}_m$ versus scaled temperature $ \delta_0(T) \equiv (T- T_d) / T_d$ for $E_0=3$, and various $A_1$ elastic constants for the four transitions. (The  symbol colours  top to bottom denote  transitions in the same  order as Fig 5.)  On approaching $T_d$, there are increasing standard-deviation  error bars, suggesting a  broadening rate distribution. }
\end{center}
\end{figure}

\begin{figure}[ht]
\begin{center}
\includegraphics[height=7.0cm, width=8.5cm]{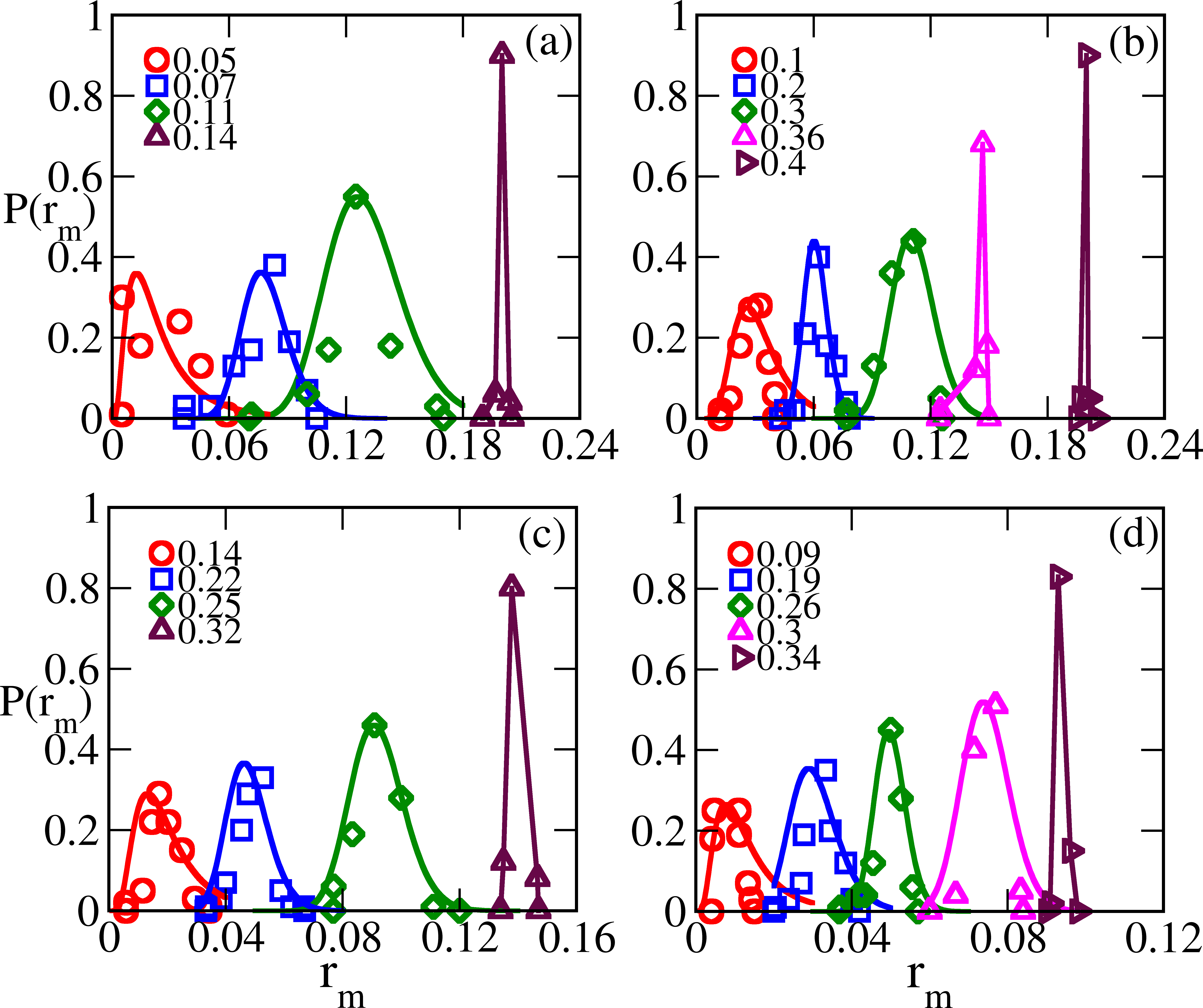}
\caption{{\it Distribution of conversion rates:} Four-panel plot of conversion rate distributions $P(r_m)$ versus  rate $r_m$ for $E_0 =3, A_1 =4$ and  different $|\delta_0 (T)|$ as in the legend.   Log-normal distributions are shown as guides to the eye. Transitions are again in the order a) tetragonal-orthorhombic transition; b) cubic-tetragonal transition; c) cubic-orthorhombic transition; and d) cubic-trigonal transition.  Fast processes are narrow, while slow processes are broad. }
\end{center}
\end{figure}

The Landau  free energy  functionals $f_L(\vec e)$  for a first order transition can be scaled to be independent or weakly dependent,  on material parameters \cite{R21}. With  $N_{OP}$  physical strains as a vector $\vec e$  in OP space, the Landau free energy  $f_L(\vec e)$ in the austenite phase  always has  a turning point at $\vec e = \vec 0$. In the martensite phase, it additionally develops $N_V$ variant minima at  $\vec e \neq 0$. 

For first-order transitions, the scaled temperature   is  \cite{R21,R22,R23,R24,R25,R26},

\begin{equation}
 \tau(T)\equiv (T-T_c)/(T_0-T_c).
\end{equation}
All energies are scaled in the thermodynamic Landau temperature where austenite and martensite free energies cross, so the scaled  $T_0 =1$. Here $T_c < T_0$ is the spinodal temperature where the austenite minimum vanishes, so uniform bulk austenite becomes unstable for $T< T_c$. 
 
The local vector OP can be written as a product of magnitude and direction $\vec e(\vec r)  = |\vec e (\vec r)|{ \vec  S}(\vec r)$. The $N_V$ directions of variant or `spin' vectors ${\vec S} (\vec r)$ identify  the degenerate variants on either side of a Domain Wall (DW), with all having unit magnitude, $\vec S (\vec r) ^2 = \sum_\ell S_\ell (\vec r) ^2 = 1$. Since austenite is always a Landau turning point, and  in any case austenite could be induced at any $T$  by local stresses, we  always  also include the austenite origin point  $\vec S(\vec r) = \vec 0$ as an allowed value \cite{R21,R22,R24}.
  
The strain magnitudes  are flat, deep into domains on either side of narrow Domain Walls that are zeros of the OP. The  local strain  magnitude is set equal to the uniform Landau mean value  \cite{R21},  $|\vec e (\vec r)| \simeq   {\bar \varepsilon} (T) > 0$, so components $\ell =1,2..N_V$ are approximated as 

\begin{equation}
e_\ell (\vec r ) \rightarrow {\bar  \varepsilon} (T) S_\ell (\vec r) .~~
\end{equation}
\noindent Substituting into the variational  free energy density \cite{R21}  with   Landau,Ginzburg, and Compatibility terms  $f= f_L (\vec e) + f_G (\nabla \vec e) + f_C(\vec  e)$, generates  a $T$-dependent effective  spin  Hamiltonian $H(\vec S, T)$, with the same three terms, inheriting material-specific parameters such as $T_c,T_0, A_1$. Each  of the  discretized-strain clock-like  Hamiltonians  have been  systematically derived \cite{R21} from continuous-strain free energies. They are bilinear in the spins, and encode the crystal symmetries, strain nonlinearities, and Compatibility constraints.

 The DW  Hamiltonian, with $E_0$ an energy scale (in units of $T_0$), is
\begin{equation}
\begin{array}{rr}
\displaystyle{F = E_0 \sum_{\vec r}[f_L + f_G + f_C ] }\\
 \displaystyle { \rightarrow H(\vec S, T)= H_L(\vec S, T) + H_G (\nabla \vec S) + H_C (\vec S)  .}
\end{array}
\end{equation}

Notice $H$ has an  inherent  separation of time scales, with the magnitude  ${\bar \varepsilon} (T)$ responding  immediately  to quenches $T < T_0$ in a single  time-step, while the   more sluggish DW adjustments  of $\vec S (\vec r)$ can take hundreds or thousands of  MC time-steps.

With  $\vec S ^6 = \vec S^4 = \vec S^2 =0,1$, the Landau term is 
\begin{equation}
\sum_{\vec r} f_L ( \vec e) \rightarrow \sum_{\vec r} f_L ( {\bar \varepsilon} \vec S) =  f_L(T) \sum_{\vec r} {\vec S} (\vec r) ^2 . ~~~
 \end{equation}
 The Landau  free energy density is

\begin{equation}
f_L (T) \equiv {{\bar \varepsilon} (\tau)}^2 g_L (T) \leq 0,
\end{equation}
defining a factor $g_L(T) $, that  vanishes at the Landau  transition  temperature $g_L (T_0) =0$, and  is negative below it.

 For  a {\it uniform} variant  $\vec S (\vec r) = \vec S_0$ a constant vector,  or in Fourier space $\vec S (\vec k) \sim \delta_{\vec k, 0} \vec S_0$, there is a vanishing of the Ginzburg term $\sim {\vec k}^2$,  and of the  Compatibility kernel $ \sim \nu _{\vec k} = 1- \delta _{\vec k, 0 }$. The uniform (Landau) free energy then sets a lower bound to the energy,
$H (\vec S, T)  \geq  N n_m (t) f_L (T) < 0$, where the martensite fraction is 
\begin{equation}
n_m (t) \equiv (1/N)  \sum_{\vec r} {\vec S (\vec r,t)}^2.
\end{equation}
so $n_m =1$  or $0$ for  uniform martensite or austenite. 

As a 2D illustration \cite{R24,R25,R26}, the square-rectangle  OP is  a scalar,  so $N_{OP} =1$. There are  two variants (rectangles along either $x$ or $y$ axes), so $N_V =2$.
The Landau free energy  $f_L = e^2 [ (\tau -1) + (e^2 -1)^2 ]$ is a triple well in the OP strains. For $\tau =1$, the three well depths at $e = 0, \pm 1$ are degenerate at zero.  For $0 < \tau < 1$ the austenite well at $e=0$ is metastable, and goes unstable at  $\tau =0$, the $T=T_c$ spinodal temperature. 
The Hamiltonian is diagonal in $\vec k$ space, $\beta H = (D_0 /2) \sum_{\vec k} [\epsilon (\vec k, T) |S(\vec k)|^2]$, where $D_0 \equiv  ( 2{\bar \varepsilon}^2 E_0  /T) $.  The energy spectrum  for long wavelengths  is $\epsilon (\vec k, T) = [- | g_L (T) |+{ \xi_0}^2 k^2 + A_1 U( \hat k)]$. The square-rectangle transition  kernel depends on direction  $\hat k = {\vec k} / |\vec k|$,  or  the single polar angle $\phi$, as 

\begin{equation}
 U (\vec k) =\frac{ \nu_{\vec k} ({\hat k}_x ^2 - {\hat k}_y ^2 )^2}{ [ 1 + (8 A_1 /A_3) {\hat k}_x ^2 {\hat  k}_y ^2 ]}=\frac{ \nu_{\vec k} (\cos 2 \phi)^2}{[1+(2A_1/A_3) (\sin 2\phi)^2]}
\end{equation}
where $A_3/A_1$ is the ratio of a non-OP (shear) elastic constant, and the non-OP compressional $A_1$. The (positive) kernel  has a  maximum value $U(max) =1$ at $\phi= \pm \pi/2$, and a minimum  value  $U (min) =0$ at  $\phi = \pm \pi /4$, driving a preferred DW orientation along both diagonals.

The energy spectrum for $A_1=0$ is a parabola pulled down to negative values by the Landau term, $\epsilon \sim[ k^2 - |g_L (T)|]$. A zero energy contour in  $(k_x, k_y)$ space is a circle with a T-dependent radius $\sqrt{|g_L(T)|}$, that shrinks to a point at $T_0$. For $A_1 \neq 0$, the bottleneck  becomes angularly modulated, with a squared-radius $k^2 (T,\phi) = |g_L (T)| -(A_1/2) U (\hat k)$, interpolating between a $\phi =\pm \pi/4$ outer radius  $k^2_{outer}(T) = |g_L (T)|$ and a $\phi = \pm \pi/2$  inner radius $k^2_{inner}(T) = |g_L (T)| - (A_1/2)$. The inner radius clearly vanishes at some temperature  $|g_L(T_d)| = (A_1/2)$ where $T_d < T_0$. This characteristic temperature, from an interplay between Landau, Ginzburg, and Compatibility terms, is  where the entropy barrier diverges.

Planes et al  \cite{R36} consider a uniform-martensite model with a Landau variational term $f_L (e,T)$.  Fast or slow behaviour is through first-passage-time jumps crossing an energy barrier, that collapses at $T_c$, or is largest at $T_0$. Our spatially non-uniform martensite model with Ginzburg, Compatibility and Landau variational terms,   differs in detail,  but is similar in spirit. Fast or slow behaviour is through MC searches crossing an entropy  barrier, that collapses at $T_1$ and diverges at $T_d$.

\subsection{  DW Hamiltonians for four transitions :}

Clock models have  discrete spins directed at  $N_V$ points on a unit circle, and are denoted by $\mathbb{Z}_{N_{V}}$, where the Ising model is  $\mathbb{Z}_{2}$. Here we generalize to  include $\vec S =\vec 0$, and call these   `clock-zero' models, denoted by $\mathbb{Z}_{N_{V}+1}$.

Drawing on Equ (7), the generic coordinate-space Hamiltonian is 

\begin{equation}
\begin{array}{rr}
\beta H = \dfrac{D_0}{2} [ \sum_{\vec{r}} \sum_{\ell} \{g_L (T) {\vec S_\ell}(\vec{r})^2 + \xi_0 ^2 |\vec \nabla {\vec S}_\ell(\vec{r})| ^2\} \\ + 
\dfrac{ A_1}{2} \sum_{\vec{r},\vec{r}'}\sum_{\ell,\ell'} U_{\ell\ell'}(\vec{r}-\vec{r}') {\vec S}_{\ell}(\vec{r}) {\vec S}_{\ell'}(\vec{r}' )],
\end{array}
\end{equation}
where the overall energy scale is $D_{0} (T) \equiv 2 {\bar{\varepsilon}^2} (T) E_0 /{ T}$. 
Here $\xi_0$ is the domain-wall thickness parameter, $A_1$ is the elastic constant for the non-OP compressional strain   \cite{R24,R25,R26}. The kernel   $U_{\ell \ell'}$ is an $N_{OP} \times N_{OP}$ matrix potential, that carries the spatial dimensionality $d=3$, and depends on  ratios of other non-OP elastic constants to $A_1$. {\it Local} meanfield treatments\cite{R22}  yield even the  complex strain textures of some real materials \cite{R13,R14}.

The generic $\vec k$ space  Hamiltonian is obtained from $S_\ell (\vec r) = \frac{1}{\sqrt N} \sum_{\vec k} S_\ell (\vec k) e^{i \vec k. \vec r}$, and  as ${\vec S} (\vec r)$ is  real, ${\vec S}({\vec k} )^{*} = {\vec S}(- {\vec k})$. 
The Hamiltonian and energy-spectrum are
\begin{equation}
\begin{array}{rr}
\displaystyle{\beta H = \frac{D_0}{2} \sum_{\vec k} \sum_{ \ell, \ell'}  \epsilon_{ \ell \ell'} (\vec k,T) S_\ell (\vec k) S_{\ell'} (\vec k)^* };\\
 \displaystyle { \epsilon_{\ell, \ell'} (\vec k, T) \equiv [\{ g_L(T) +  \xi_0 ^2 { \vec K}^2  \} \delta_{\ell, \ell'} + \frac{A_1}{2} U_{\ell  \ell'} ({\vec k} )].}
\end{array}
\end{equation}

\begin{figure}[ht]
\begin{center}
\includegraphics[height=7.0cm, width=8.5cm]{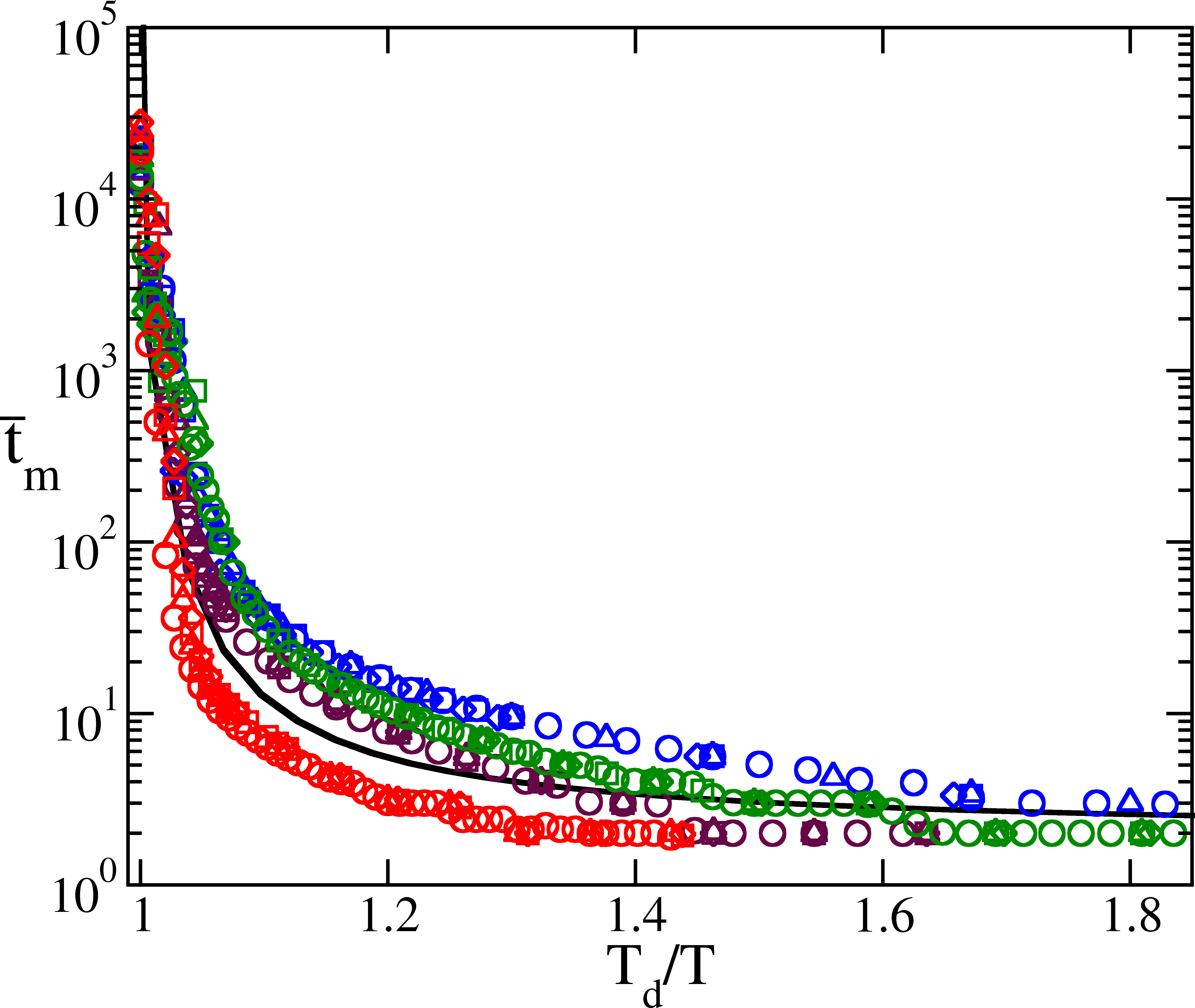}
\caption{{\it  Angell-type plot as in glasses,  for athermal martensites:  } The delay times ${\bar t}_m (T)$ are replotted versus $T_d/T$, so that Arrhenius  activated times would give a straight line. The x-axis data for the four transitions go from  a value of unity  on the left,  to  their respective $T_d/T$ on the right. The  obvious  curvature shows that delays are not activated over energy barriers. They are attributed to {\it entropy} barriers.}
\end{center},
\end{figure}

The 3D transitions  are given below in the sequence of  $N_{OP} =1,2,2, 3$, plus $N_V=2,3,6,4$  cases respectively.\\

\noindent {\it a) Tetragonal-Orthorhombic case ($N_{OP} =1, N_V=2$):}\\
The  scalar OP is  the  first deviatoric-strain OP $e_2 = (e_{xx} -e_{yy}) / {\sqrt 2}$, with  $\vec S$ having $2 + 1$ values: at the origin, and pointing to the two endpoints of a unit-circle diameter. The Hamiltonian is like a 3D Spin-1 Blume Capel model \cite{R24}, but with anisotropic powerlaw interactions, and  with the quadratic term $ f_L(T) \vec S^2$ where $f_L (T) <0 $. 
The Hamiltonian is a clock-zero $\mathbb{Z}_{2+1}$ model \cite{R21}.The scalar compatibility kernel $U(\vec k)$ for the tetragonal-orthorhombic transition  is given in Equ. (A26) of Ref 21.  

 With  $f_L(T) \equiv {\bar \varepsilon}^2(T) g_L (T)  \leq 0$, the Tetragonal-orthorhombic  (and  also 2D square-rectangle) case has the squared-mean OP and $g_L(T)$ factor as \cite{R21} 

\begin{equation}
\begin{array}{rr}
\displaystyle{{\bar \varepsilon}^2(T)=(2/3)[1+\sqrt(1-3\tau/4)]};\\
\displaystyle{g_L (T) \equiv =(\tau-1)+({\bar \varepsilon}^2 -1)^2    }.
\end{array}
\end{equation}

\noindent { \it b) Cubic-tetragonal  case ($N_{OP} =2, N_V =3$):}\\
This  3D transition has been considered earlier\cite{R23,R31}.
The OP strains are the two deviatoric strains $(e_3, e_2) = (\{2 e_{zz} - e_{xx} - e_{yy}\} /{\sqrt 6}, \{e_{xx} -e_{yy}\}/{\sqrt 2})  $.
The spin values  $\vec S =(S_3, S_2)$  vectors  are in a plane in OP space, 
 with  $\vec S$ having $3 + 1$ values:  at the origin and pointing to the three corners of a triangle inscribed in a unit circle.
The Hamiltonian is a clock-zero $\mathbb{Z}_{3+1}$ model \cite{R21}.The  compatibility kernel  is a $2\times 2$ matrix, $U_{\ell  \ell'} ({\vec k} )$, with $\ell, \ell' =2,3$, as in Equ (A23) of Ref 21.

The  mean OP and   $g_L(T)$   of the  cubic-tetragonal transition  are 
\begin{equation}
\begin{array}{rr}

\displaystyle { {\bar {\varepsilon}}(T)=(3/4)[1+\sqrt(1-8\tau/9)] };\\
\displaystyle{g_L (T)  =(\tau-1)+({\bar \varepsilon}-1)^2    }.
\end{array}
\end{equation}

\noindent {\it  c) Cubic-orthorhombic case ($N_{OP} =2, N_V =6$): }\\
The OP strains  are again  the two deviatoric strains $(e_3, e_2)$ as above. 
The nonzero  $\vec S =(S_3, S_2)$ spin vectors are  in a plane in OP space, 
with  $\vec S$ having $6 + 1$ values: at the origin and pointing to the six corners of a hexagon inscribed in a unit circle. 
The Hamiltonian is a clock-zero $\mathbb{Z}_{6+1}$ model \cite{R21}. The Compatibility  kernel is a $2 \times 2$ matrix ${U_{\ell, \ell '}} (\vec k)$,  again  with $\ell, \ell' = 2,3$ and is the same as the cubic-tetragonal case, given  in Equ (A23) of Ref 21. 

The squared-mean OP and $g_L (T)$  for the Cubic-orthorhombic case, are the same as the tetragonal-orthorhombic in  Equ (15).

\noindent  { \it d) Cubic-trigonal  case ($N_{OP} =3, N_V =4$):}\\
 The three OP for the cubic-trigonal transition are the three shears $e_4,e_5, e_6 = e_{xy}, e_{yz}, e_{zx}$, and the variant vector has 
 three vector components, 
with  $\vec S$ having $4 + 1$ values: at the origin and pointing to the four corners of a tetrahedron inscribed in a unit sphere. The Compatibility kernel  is now a $3 \times 3$  matrix $U_{\ell, \ell'} (\vec k)$, with $\ell, \ell' =4,5,6$, or  six components $U_{44}, U_{55}, U_{66}, U_{45}, U_{54}, U_{64}$  as in Equ. (A20) of Ref 21.
 
The mean OP and $g_L (T)$ for the cubic-trigonal case are the same as the cubic-tetragonal case in Equ (16).  
  
\section{ Monte Carlo Simulations:}

Simulations were done on  models for four transitions  in 3D. The initial state $t=0$ is  $2\%$ randomly and dilutely seeded martensite cells, in an austenite sea of  $\vec S =0$.        Evolutions proceed at  quench temperatures $T < T_0$.
Typical parameters are $T_0=1$; $\xi_0 ^2=1$; $T_c=0.81$ to $0.97$; $A_1=1$ to $85$; $E_0=3,4,5,6$; $N= L^3=16^3$; $N_{runs}=100$; and holding times $t_h=10^4$ MC sweeps,  each over $N$ sites.

The Compatibility kernels arise from the non-OP harmonic terms, with $(6-N_{OP})$  elastic constants.  For all transitions, we specify the fixed ratio of  other non-OP elastic constants, to   $A_1$. For the tetragonal-orthorhombic transition, with  $N_{OP} =1$  and $S_2$ as the OP spin,  the  non-OP elastic constants  are the other deviatoric constant $A_3$, and the three shear constants $A_4, A_5, A_6$, set to be $A_3 = A_4 = A_5 = A_6 = A_1 /2$. Similarly, for the cubic-tetragonal and cubic-orthorhombic cases with  $N_{OP} =2$ and  $(S_3, S_2)$ as the OP spins,  the non-OP constants  are set to be
$A_4 = A_5 = A_6 = A_1 /2$. Finally, for  the cubic-trigonal  case  with $N_{OP}=3$  and the shears  $(S_4,S_5,S_6)$ as the OP,  the constants are set as   $A_2 = A_3 = A_1 /2$.

 The standard MC procedure \cite{R32} is followed, with an extra data retention \cite{R1,R2,R3,R4} of energy changes.  \\
0. Take $N$ lattice sites, each with a  vector spin of $N_{OP}$ components,  in one of $N_V +1$ possible values (including zero) at MC time $t$. Each $\{{\vec S}(\vec r)\}$ set is a `configuration'. With $n_m=0$ or $1$ for uniform austenite or martensite, the average martensite fraction is $n_m(t)  \le 1$.   The  conversion time $t_m$  is defined as when \cite{R24}   $n_m (t_m) =1/2$.
 \\
1. Randomly pick one of  $N$ sites, and randomly flip the spin  on it  to a {\it new} direction/value, and find  the (positive or negative)  energy change $\delta E$.\\
2. If $\delta E \leq 0$, then accept the flip. If $\delta E > 0$, then accept flip with probability $e^{- \delta E /T }$. {\it Record} this  spin-flip $\delta E$, that is usually not retained after use.\\
3. Repeat steps 1 and 2. Stop after $N$ such spin-flips. This is the $t+1$ configuration with  $n_m (t+1)$. \\
4. We collect  all  recorded $ \{ \delta E\}$ over each MC sweep of every run while tracking $n_m(t)$. The collection is  only  up to a waiting time  equal to the martensite conversion  time.  $t=t_w \leq t_m (T) \leq t_h.$                          
We do six quenches, from $T=T_1$ upwards to $T_d$. 

 A single-variant  athermal martensite droplet or `embryo'  \cite{R11} can form  anywhere, and after a local conversion at a waiting time  $t_w=t_m$, can  {\it propagate} rapidly to the rest of the system \cite{R11,R19}. 
Hence it is the mean  {\it rates} ${\bar r}_m$ (or inverse times),  that are averaged over $N_{run}$ runs, analogous to total  resistors in parallel  determined by the smallest resistance. The mean times ${\bar t}_m$ are defined as the inverse mean 
rates\cite{R24}.
  
\begin{equation}
{\bar r}_m (T) =\frac{1}{ N_{run}}  {\sum_{n =1}}^{N_{run}} \frac{1}{t_m (n)} ; ~~{\bar t}_m (T) \equiv \frac{1}{{\bar r}_m (T)}.
\end{equation}
 
 For  an un-successful $n$-th run that  does not convert in a holding  time $t_h$, the rate $r_m (n)$ could be  assigned a value of  either  $1/t_h$ (conversion right after cutoff) or $0$ (conversion never occurs). 
  We choose the $1/ t_h$ cutoff, that shows  up as a flattening of the mean rate near $T_d$.  An extrapolation  of the linear part of $r_m(T)$ to the temperature axis yields a value\cite{R24} for $T_d$.

 \begin{figure}[ht]
\begin{center}
\includegraphics[height=7.0cm, width=8.5cm]{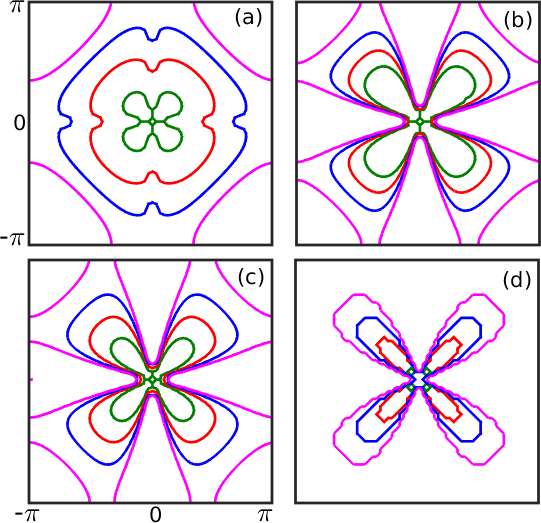}
\caption{{\it Bottlenecks in phase space  for different transitions:} Four-panel plot shows energy contours on 2D  slices of the 3D bottleneck, with  $\delta_0 (T)= -0.1, 0, +0.1, +0.2$. The sizes decrease on warming, and the  bottleneck inner radius is seen to pinch off, at some ${T_d}'$. The contours  are a) Tetragonal-orthorhombic, ${T_{d}}' = 0.69$ b) Cubic-tetragonal, ${T_{d}}' = 0.60$; c) Cubic-orthorhombic  ${T_{d}}' =0.60$; d) Cubic-trigonal ${T_{d}}' = 0.70$. The corresponding delay-divergence temperatures $T_d =0.75, 0.59, 0.69,0.75$ are close in value, and taken as the same, for simplicity of discussion.}
\end{center} 
\end{figure}

\begin{figure}[ht]
\begin{center}
\includegraphics[height=7.0cm, width=8.5cm]{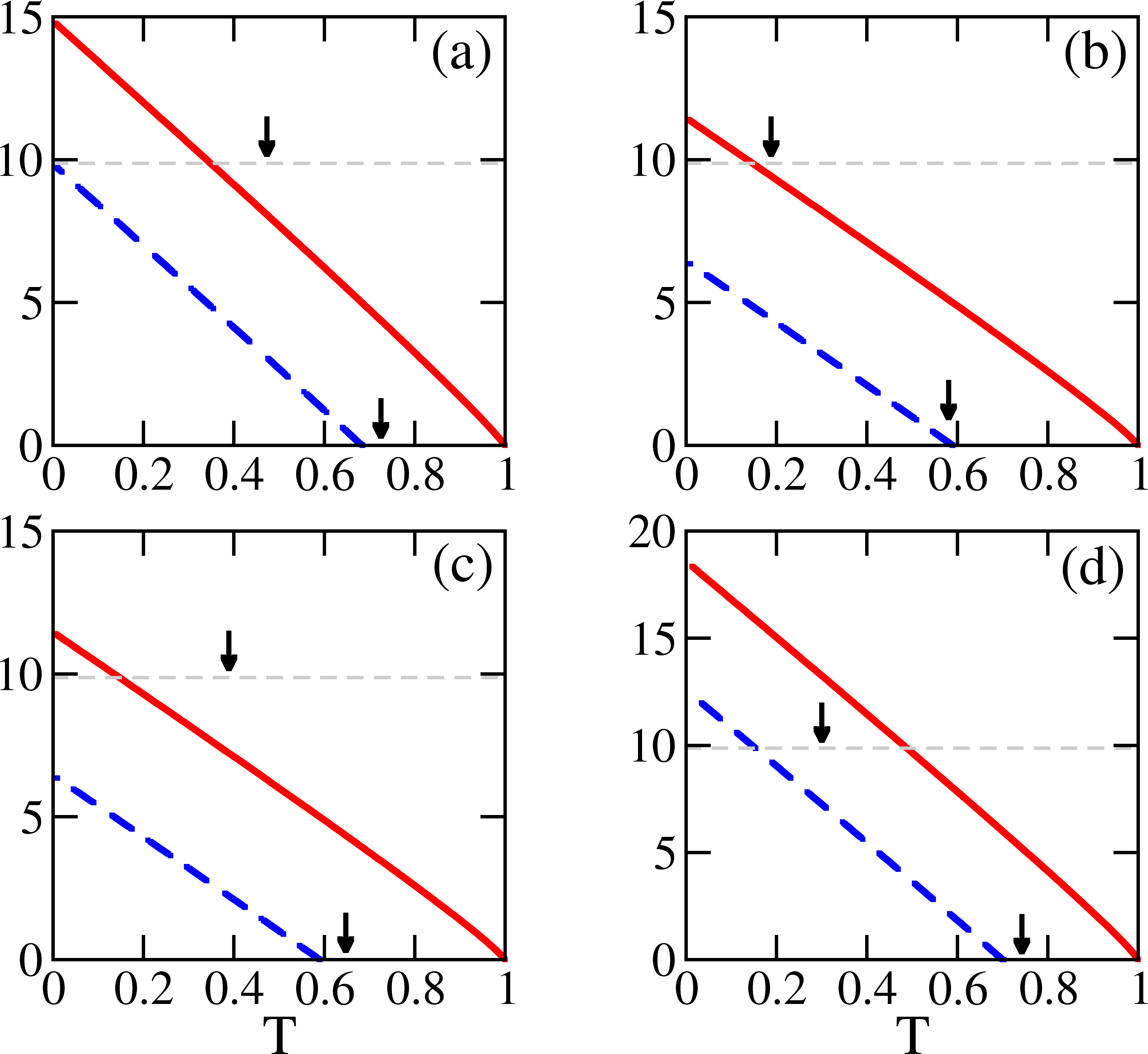}
\caption{{\it Bottleneck inner and outer radius: }  Four-panel plot showing $k_{outer}^2 (T)$ (solid), $k_{inner}^2 (T) $ (dashed) lines versus temperature $T$, for transitions in the order a) to d)   as in Fig 9. The inner radius vanishes close to the  delay-divergence $T_d$ (downward arrows). The outer radius vanishes at the thermodynamic transition temperature $T_0 =1$. It  intersects  the horizontal light dashed line denoting a Brillouin Zone scale $\pi^2$, at temperatures close to the $T_1= 0.51,0.21,0.4,0.31$  for explosive conversions (downward arrows)  }
\end{center}
\end{figure}

\section{Delays and bottlenecks:}
For the four transitions, we consider i) Delays from sluggish  Domain Walls; ii)  Bottlenecks in Fourier space.

\subsection{Delays from sluggish Domain Walls}
Figure 2 shows the twinned microstructures in all four transitions,  colour coded  through the variant-label $V$, where $V=0$ is always austenite. For the single-OP tetragonal-orthorhombic transition,  $V=\pm 1$ has two variant colours. For the two-component OP, the cubic-tetragonal  and cubic-orthorhombic transitions have respectively, $V=1,2,3$ and $V=1,2,...6$ variant  colours.  For the three-component OP, the cubic trigonal has $V=1,2,3,4$ four variant colours. 
 Fig 2 shows that  {\it all}  allowed  degenerate $N_V$  variants are  present, for all four transitions. 

The tetragonal-orthorhombic, and cubic-tetragonal twins can have Domain Walls decorated with austenite, as also found in the 2D case \cite{R25}.  Such observed  austenite retentions\cite{R11}  can be understood: they are  energetically favoured, when the  lower Ginzburg costs of  austenite-martensite walls compensate for the absence of the negative condensation energy $f_L (T)<0$ of martensite unit cells.  As $T$ is lowered, the energy accounting is reversed,  and the austenite inclusions are expelled, replaced by martensite, so DW are only between martensite variants \cite{R25}.

We define a fractional deviation from a characteristic temperature $T_d$, as
\begin{equation}
\delta _0 (T) \equiv (T- T_d) / T_d  \leq 0.
\end{equation}

Figure 3 shows the martensite conversion-fraction $n_m(t)$ versus MC time $t$ in a {\it single run},  for different $\delta_0 (T)$.  
For quenches to below $T =T_1$,  there is an immediate avalanche conversion in a single $t=1$ sweep, characteristic of   athermal martensite. 
For temperatures $T > T_1$  there is a strange `incubation' behaviour, or a postponement of these  avalanches. The fraction $n_m(t)$  remains virtually unchanged, up till $t = t_m$  when $n_m (t)$ rises sharply through $1/2$,  to  unity. The  cubic-trigonal transition has $N_{OP} =3$ order parameter components, $N_V =4$ variants, and can show unusual behaviour.  Here, there is an  initial  jump to $n_m =1/2$  followed by incubation, and  then a jump to unity. 
 
Fig 4 shows that for $T< T_1$  there is an immediate spike in the  MC acceptance fraction $A_{acc} (t)$  at $t=1$, and drop in  energy  $E(t)$ to negative values. For $T> T_1$, during  incubations they both remain zero,  up to  $t= t_m (T)$, when the acceptance spikes and the energy drops. Again, the cubic-trigonal case is unusual.

What goes on  microscopically, during incubation ? In 2D,  Video \cite{R27} A shows random initial seeds of both variants (red,blue)  can quickly form an almost zero energy   single-variant martensitic droplet  or embryo (red)  in an austenite  background (green), in the DW Vapour state. The small droplet extends and retracts amoeba-like arms, searching for energy-lowering  pathways. After a  long conversion delay $t_m$ of hundreds of time steps, when $n_m \simeq 0$,  the single-variant droplet (red) suddenly expands rapidly and generates the opposite variant  (blue). This is the wandering-wall or DW Liquid state. After a shorter orientation delay, the walls of the DW  Liquid orient to a  DW crystal.

Figure 5 shows for all four transitions,  the mean fraction of successful conversions $\Phi_m$  during a holding time $t=t_h$ over 100 runs, versus the temperature deviation $\delta_0 (T)$. For temperatures  $\delta_0 \leq -0.1$, every run converts,  and $\Phi_m =1$.  However,  for $\delta_0 > -0.1$, $\Phi_m (T)$ falls  through  1/2  at $\delta_0 (T) \sim -0.05$, and then  to zero.  The success fraction  is not exponentially sensitive  to overall energy scales $E_0=3,4,5,6$, so the probability of conversion is not activated over an energy barrier. 

Just above  $T_d$ there are unsuccessful runs (not shown), when  the martensite seeds can dissolve back into austenite. The seeds will not be regenerated, even if the holding time $t_h$ is increased, or if the temperature is lowered\cite{R20}.

Figure 6 shows  a linear-linear plot of the mean conversion times  ${\bar t}_m$ versus  $\delta_0 (T)$ for various $E_0$. The absence of exponential  sensitivity to $E_0$ implies the delays are  not activated over energy barriers, but are due to {\it entropy barriers}.   For $T <T_1$ fluctuations   are small, while for $T> T_1$  standard-deviation  error bars  $\pm \sigma$ over the  $N_{run} =100$ runs,   are larger,  on approaching $T_d$. 

If  the probability of entropy-barrier crossings  coming from a product of sequential random steps, a {\it logarithm} of rates would be an additive random variable, suggesting a  log-normal distribution of rates \cite{R25} $P(r_m)$. Fig 7 shows optimized-bin histograms \cite{R37} of  $r_m$ data,  with (asymmetric) log-normal lines as guides to the eye.  Although  data are too sparse to decide distributions, clearly fast conversions are narrow, and slow conversions are broad,  similar to protein folding \cite{R29,R30}.

 Fig 8 shows an Angell-type plot \cite{R8} of log delay time versus  $T_d /T >1$.   Arrhenius-type activations over energy barriers would be linear. The delays have curvatures, and so must be from entropy barriers.The solid curve is VFT behaviour of Equ (29) below,  with $B_0=0.25$.

\subsection{Bottlenecks  in  Fourier-space:}

 We draw on  concepts of protein folding \cite{R28,R29,R30}, to understand the entropy barrier delays.

A purely random search of protein configurations could take astronomical times (Leventhal paradox). Rapid protein folding is attributed to a  configuration-space Golf Hole (GH) opening into  a Funnel of negative-energy configurations,  leading rapidly to the folded protein state \cite{R28,R29}.   Bicout and Szabo \cite{R30} consider a random walk of a Brownian particle in a space of eigen-labels of protein  folding modes (that could be analogous to propagative martensitic  twinning modes\cite{R38}). The Brownian particle has to locate and enter spherical zero-energy GH contour of marginal modes. Unusual delays can occur, at the GH  edge.

  \begin{figure}[ht]
\begin{center}
\includegraphics[height=6.5cm, width=8.0cm]{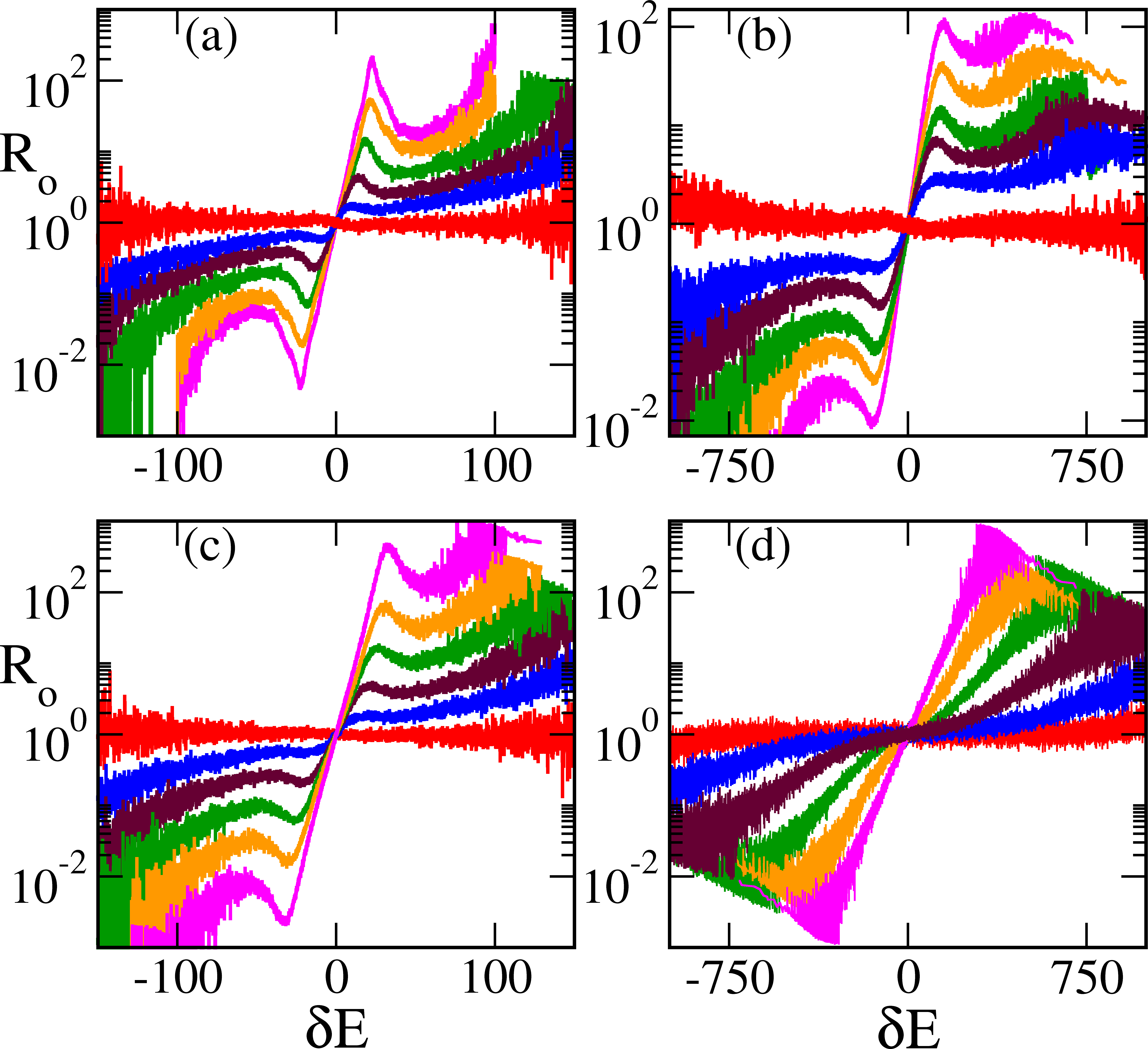}
\caption{{\it Fluctuation ratios $R_0$   of forward/ backward energy change probabilities  in log-linear plot:}   Four-panel plot for all transitions,  of  the ratio $ R_0(\delta E,T) \equiv P_0 (\delta E, T)/  P_0 (-\delta E, T)$ versus  MC energy change $\delta E$, for six $T$.}
\end{center}
\end{figure}

\begin{figure}[ht]
\begin{center}
\includegraphics[height=6.5cm, width=8.0cm]{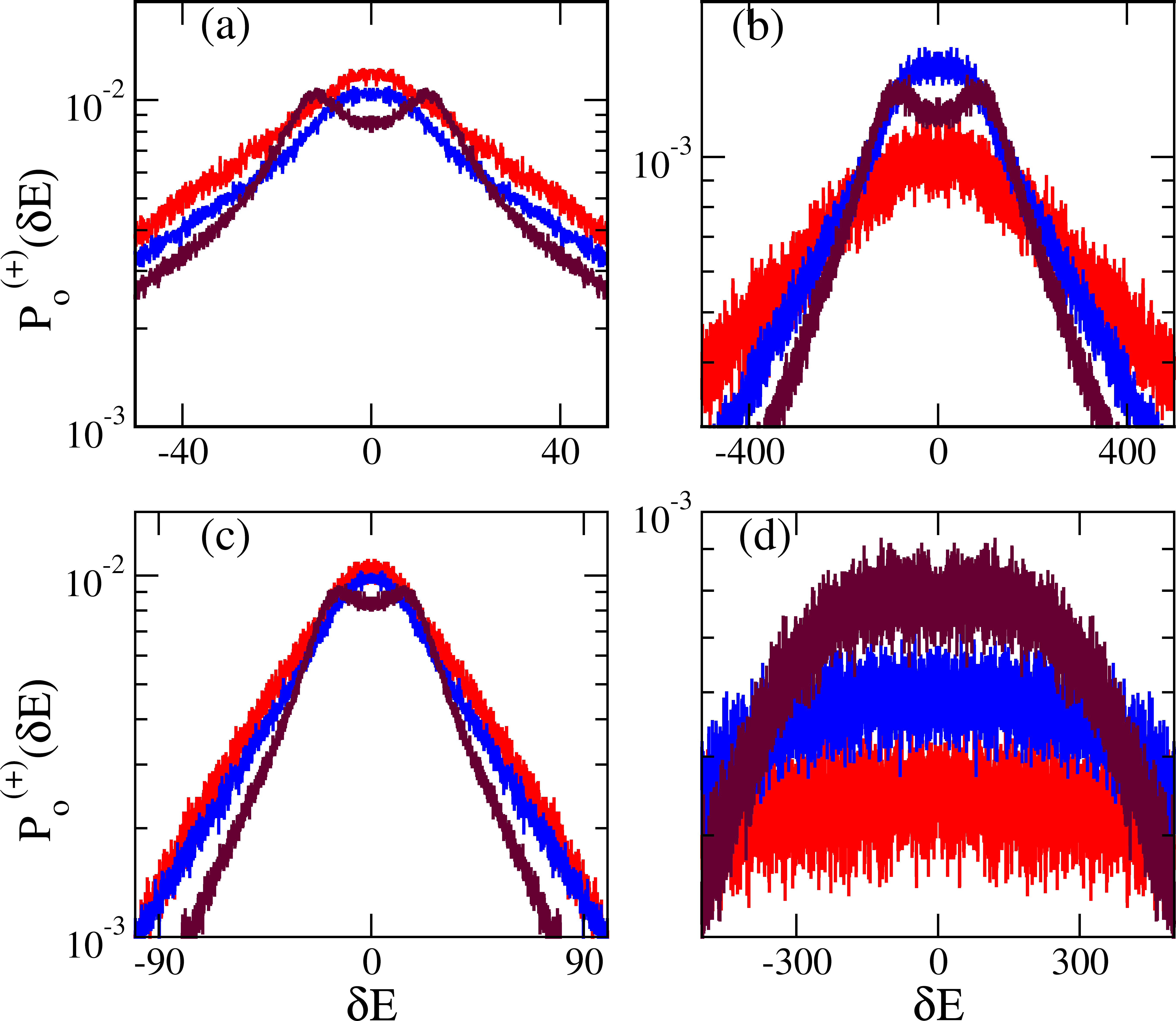}
\caption{{\it Even-symmetry  prefactor  $P^{(+)}_0$  in linear-linear plot:}  Four-panel plot for all transitions,  checking that  the  log-linear prefactor   $ P^{(+)} _0 (\delta E, T) $ versus  energy change $\delta E$  has no linear contribution near the origin. } 
\end{center}
\end{figure}

\begin{figure}[ht]
\begin{center}
\includegraphics[height=6.5cm, width=8.0cm]{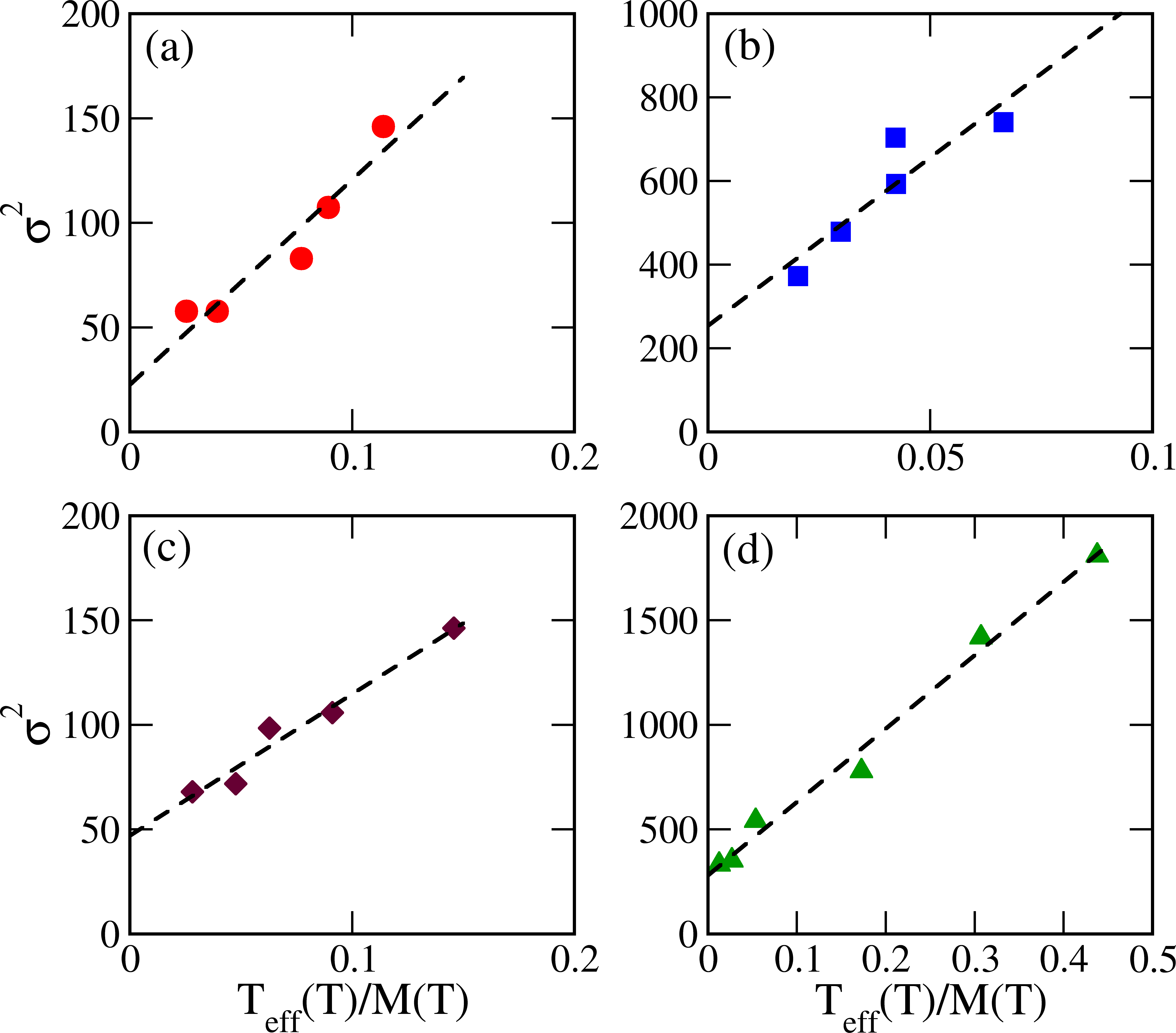}
\caption{{\it Variance of the even-symmetry prefactor :} Four-panel   linear-linear plot of the variance  $\sigma^2 (T)$,  of the prefactor $P_0 ^{(+)} (\delta E, T)$, versus  $T_{eff}(T)/ M(T)$ for the four transitions. The  energy scale is chosen as the (positive)  mean energy change  $M (T) \equiv <\delta E>$ over the entire  PES distribution.}
\end{center}
\end{figure}

In our case, the bottleneck is fixed by the energy spectrum from Equ (14)   in 3D Fourier space (taken as diagonal in $\ell$,  for discussion):

 \begin{equation}
 \epsilon_{\ell \ell}  (\vec k, T)  = \xi_0^2{{ \vec K} (\vec k)}^2 - |g_L(\tau)| + (A_1 /2) U_{\ell \ell} (\hat k).~~
\end{equation}
Energy spectra set to a constant $C$  or $\epsilon_{\ell \ell }  (\vec k, T) = C $, define contours  in $\vec k$-space.
The Ginzburg term  at long wavelengths  $\sim \vec k^2$, forms  a $\vec k$-space spherical bowl in 3D, with zero-energy minimum at  the origin.  
The Compatibility term $U_{\ell  \ell} (\hat k)$ angularly modulates its  2D surface to produce an anisotropic zero-energy  contour \cite{R24}.

The phase-space boundary  $\epsilon_{\ell \ell}  (\vec k, T) =0$  separates an outside $\vec k$ region of positive (austenite) energies, from a $\vec k$-region inside, of (martensitic) negative energies.  Video B shows  \cite{R27}   a  2D circular droplet Fourier profile in   $(k_x, k_y)$ as it distorts, to enter the phase space bottleneck.

We consider spectra with $\epsilon_{22} (\vec k, T)$ for the first three transitions, and $\epsilon_{66} (\vec k, T)$ for the cubic-trigonal case. 
Consistent with the twin orientations in Fig 1, the plane  intersecting the 3D bottleneck to yield a 2D cross-section  is taken as  $[{\hat k}_x,{\hat k}_y,{\hat  k}_z] = [1,1,1]$. 
The plane through the Brillouin Zone origin  is $k_x + k_y + k_z =0$, and  Fig 9 shows  for all four transitions,  the  $T$-dependent contours  of constant   $\epsilon_{\ell \ell}  (k_x , k_y ,k_z=  -k_x - k_y, T )$ versus   ($k_x , k_y)$ for temperature range  $\delta_0 (T)= -0.1$ to $+0.2$.As mentioned, the cubic-tetragonal and cubic-orthorhombic have the same kernel but different Landau factor $g_L (T)$, so one would expect  the second and third panels to show the same overall shapes,  but slightly different  energy contours for a given $T$:  this is indeed the case. 

 The bottleneck sizes  are large at low $T$ and small at high $T$. The contours are  angularly modulated between a smaller inner-radius wave-vector $k_{inner} (T)$ and larger outer-radius  wave-vector $k_{outer} (T)$. 
From the spectrum Equ (19) 
\begin{equation}
\begin{array}{rr}
\displaystyle{k_{outer} ^2  (T) ={ \xi_0}^{-2} [|g_L(T)| - (A_1 /2) U_{\ell \ell} (min)]}; \\
\displaystyle{k_{inner} ^2  (T) ={ \xi_0}^{-2} [|g_L(T)| - (A_1 /2) U_{\ell \ell} (max)].}
\end{array}
\end{equation}
For $ U_{\ell \ell} (min)=0 $, the outer  square-radius vanishes at the Landau temperature, $k_{outer} ^2 (T_0) = |g_L (T_0)| = 0$. Close to transition,
\begin{equation}
k_{outer} ^2 (T) \simeq b(T_0) (T-T_0)
\end{equation}
 where the Taylor expansion coefficient $b (T) \equiv -d g_L (T)/dT <0$.

With a positive  $U_{\ell,\ell} (max) >0$   there is a temperature $T_{d} < T_0$  where the inner radius can pinch off to zero, $k_{inner }^2 (T_{d})=  |g_L (T_{d})| - (A_1/2) U_{\ell, \ell} (max) =0$, while the outer radius is still nonzero. 
Near $T_{d}$, 

\begin{equation}
\begin{array}{rr}
\displaystyle{ k_{inner }^2 (T) = |g_L (T)|  - |g_L (T_{d})| } \\
\displaystyle{ \simeq b(T_{d}) ~ (T- T_{d} ]   }.\\
\end{array}
\end{equation} 
 Figure 10 shows   the  inner and outer squared-radii, both almost  linear, and vanishing  respectively  at  $T_d$ and  $T_0$.
 
 The  conversion-delay divergence comes from a pinch-off of the inner radius  $k_{inner} (T)$ of the bottleneck. The topology of a  2D slice  of the 3D  negative energy states, goes from  an open butterfly to a segmented  four-petalled flower \cite{R25}. 
 It is impossible for the  broad Fourier profile of a small droplet to  distort  at zero total energy, into four separated petal-like segments. The lower-energy states for  $T > T_d$ are thus available, but not accessible. The intershell configurational pathway closes;  the success-fraction vanishes; and  the entropy barrier diverges.

\section{PES  signatures in all four transitions }

 We now exhibit  PES signatures in the four athermal martensite transitions. Entropy barriers are insensitive to energy scales, so we consider only $E_0=3$.

\begin{figure}[ht]
\begin{center}
\includegraphics[height=6.5cm, width=8.0cm]{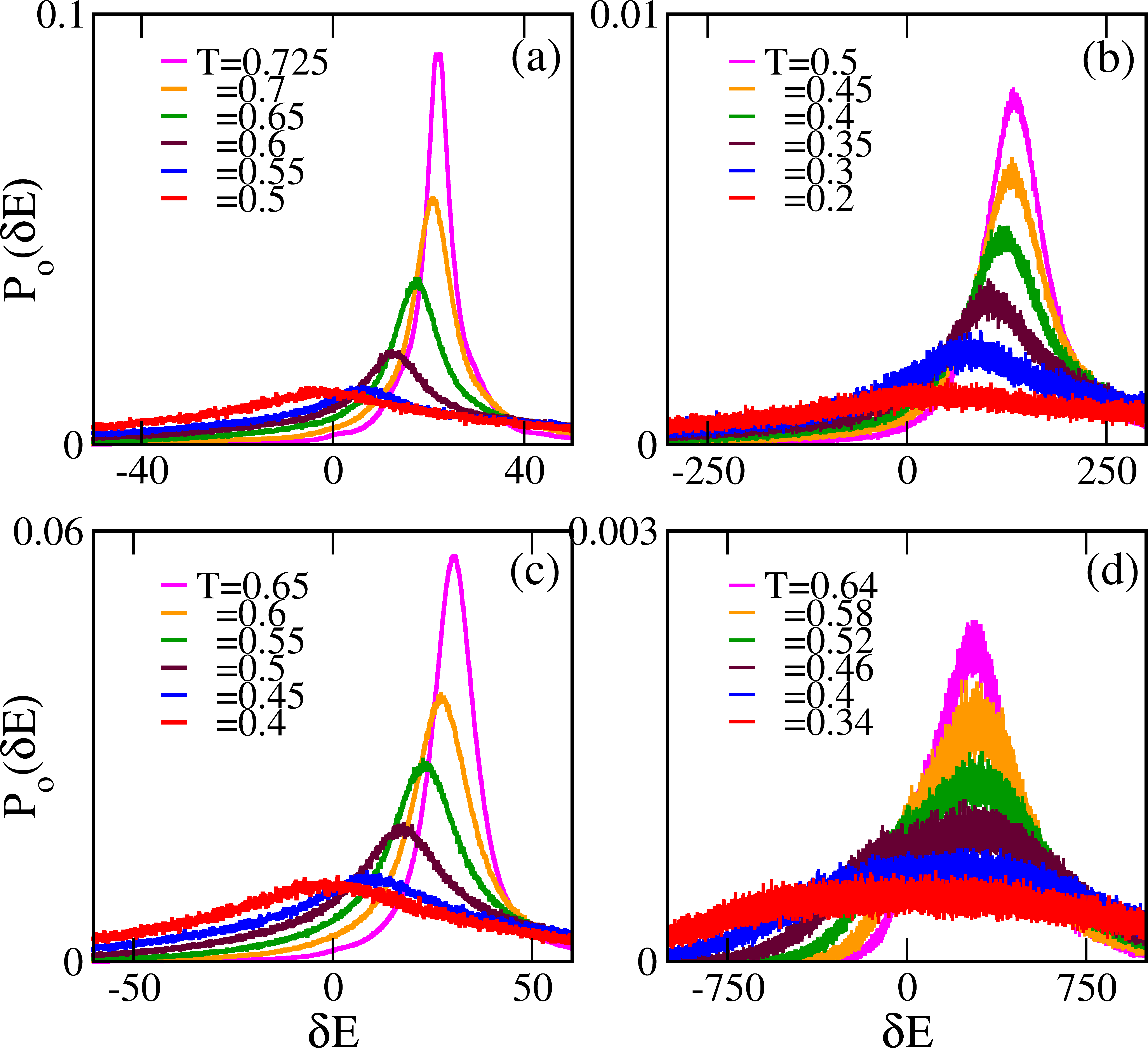}
\caption{{\it Normalized probabilities  $P_0$ of energy-changes in linear-linear plot:}  Four-panel Figure for all transitions,  showing   the probability   $ P_{0} (\delta E, T) $ versus  energy change $\delta E$  for  the six quench  temperatures $T$ in the legend.The probabilities are consistent with the predicted PES signature of  shifted Gaussians peaked near a  mean energy changes $M(T) > 0$, with exponential heat-loss tails for $\delta E < 0$. }
\end{center}
\end{figure}

\begin{figure}[ht]
\begin{center}
\includegraphics[height=6.5cm, width=8.0cm]{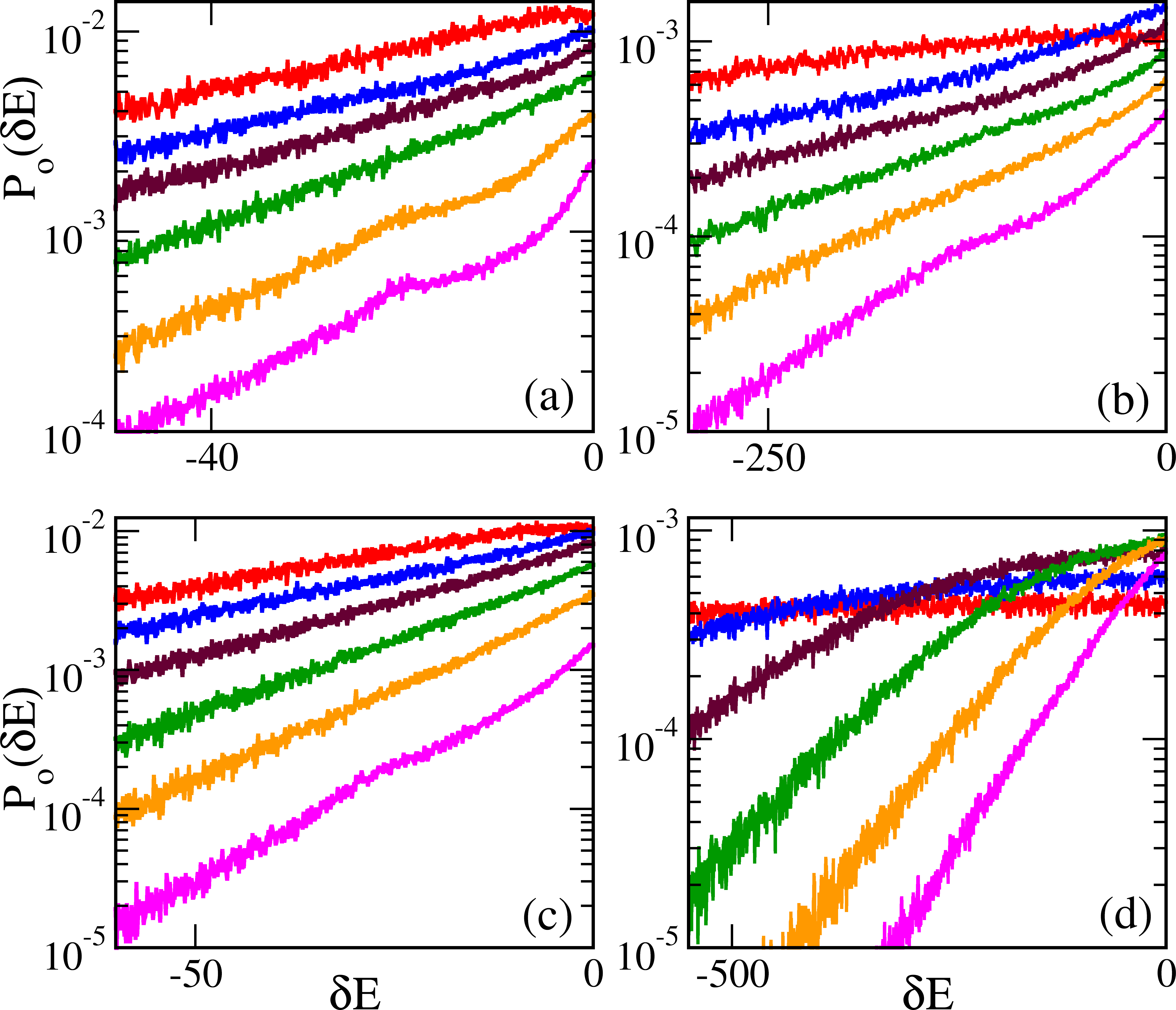}
\caption{{\it  Normalized probability  of energy-changes $P_0$  in  log-linear plot :}  Four-panel Figure for all transitions,  showing a log-linear version  of Fig 14  on a zoomed in scale  of  $ P_{0}(\delta E,T)$ versus  energy change $\delta E$, at  the  same six  temperatures $T$.The linear behaviour  is a PES signature, with  slopes determining  the inverse effective temperatures, $\beta_{eff}/2  \equiv 1/ 2T_{eff}$,   that flatten near $T=T_1$.}
\end{center}
\end{figure}
 
  As noted the MC procedures of  Section IV retain the set of  {\it single spin-flip} energy changes $\{\delta E\}$  from all N  spin-flips,  in each of $N_{run}$ runs, up to t MC  times $t \leq t_m(T)$.  The histograms can be dense since the  data set size can be large:  $N \times { t}_m \times N_{run}$ has up to  $16^3 \times 10^4 \times 100 $ data points. 
   
An ageing-state  fluctuation relation is postulated \cite{R4}. The probability $P_0 (\delta E, T)$ to  hit configurations $E'$ from $E$,  is  proportional to the {\it target} size, or the  number of accessible states:  $P_0 (\delta E,T) \sim \Omega (E')$, with $S(E') = \ln \Omega (E') $, where $E' = E + \delta E < E$.  The reverse path  has $P_0(-\delta E,T) \sim \Omega (E)$. The ratio of forward and backward probabilities $R_0(\delta E)$    is related to the entropy change and entropy barrier $\Delta S  (\delta E) \equiv S(E')-S(E) \equiv - S_B  < 0$.  Thus

\begin{equation} 
R_0 \equiv \frac{P_0 (\delta E,T)}{ P_0 (-\delta E,T)}  =\frac{ \Omega (E')}{ \Omega (E)}  = e^{\Delta S(\delta E)}.
\end{equation} 

Fig 11 shows  four-panel log-linear  plots of the fluctuation ratio  $R_{0} (\delta E,T)$.  Since $R_0(\delta E) R_0 (-\delta E) \equiv 1$, the  entropy change is odd, $\Delta S(\delta E) +\Delta S(-\delta E)=0$, and a solution is 

\begin{equation}
P_0 (\delta E,T)=  {P_0}^{(+)} (\delta E,T) ~  e^{\frac{1}{2} \Delta S(\delta E)},
\end{equation} 
 where the (even) prefactor  is the geometric mean,  ${P_0}^{(+)} (\delta E, T) = \sqrt{P_0 (\delta E,T) P_0 (-\delta E,T)}$. For small energy changes,   $S (E+\delta E) - S(E)  \simeq \beta_{eff} \delta E$,  and hence  $P_0(\delta E, T) \simeq  {P_0 }^{(+)} (0,T) ~e^{\frac{1}{2} \beta_{eff}(T) \delta E}$.
  
 Fig 12 is just a check that the prefactor $P^{(+)}_0 (\delta E, T)$ versus $\delta E$ has no linear contribution near the origin, that might modify the exponential tail. For temperatures near $T_1$  it is a  single-peak gaussian, while near $T_d$ it can go  bimodal.

 Fig 13 shows  that the variance $\sigma^2$  of the weight ${P_0}^{(+)} (\delta E,T)$ versus  a scaled $T_{eff} (T)$ is linear and nonsingular,  for all four transitions. 
 
Fig 14  shows  four-panel  linear-linear plots of  (normalized) $ P_0 (\delta E,T)$ versus $\delta E $ for four transitions, each at one of six temperatures in the Legend. The peak is at positive energy  as in the oscillator case \cite{R2}, and  moves left as $T$ decreases towards $T_1$. The exponential tails near the origin are barely  visible.
  
  Fig 15 shows the same  four plots but now in log-linear form, and zoomed in. The curves all show the PES signature of linearity around the origin  $\delta E =0$.  The cubic tetragonal panel has been shown earlier \cite{R31}.
As $T$ is lowered to  $T_1$, the slopes $\beta_{eff} (T) /2$ all flatten.   

 \begin{figure}[ht]
\begin{center}
\includegraphics[height=6.5cm, width=8.0cm]{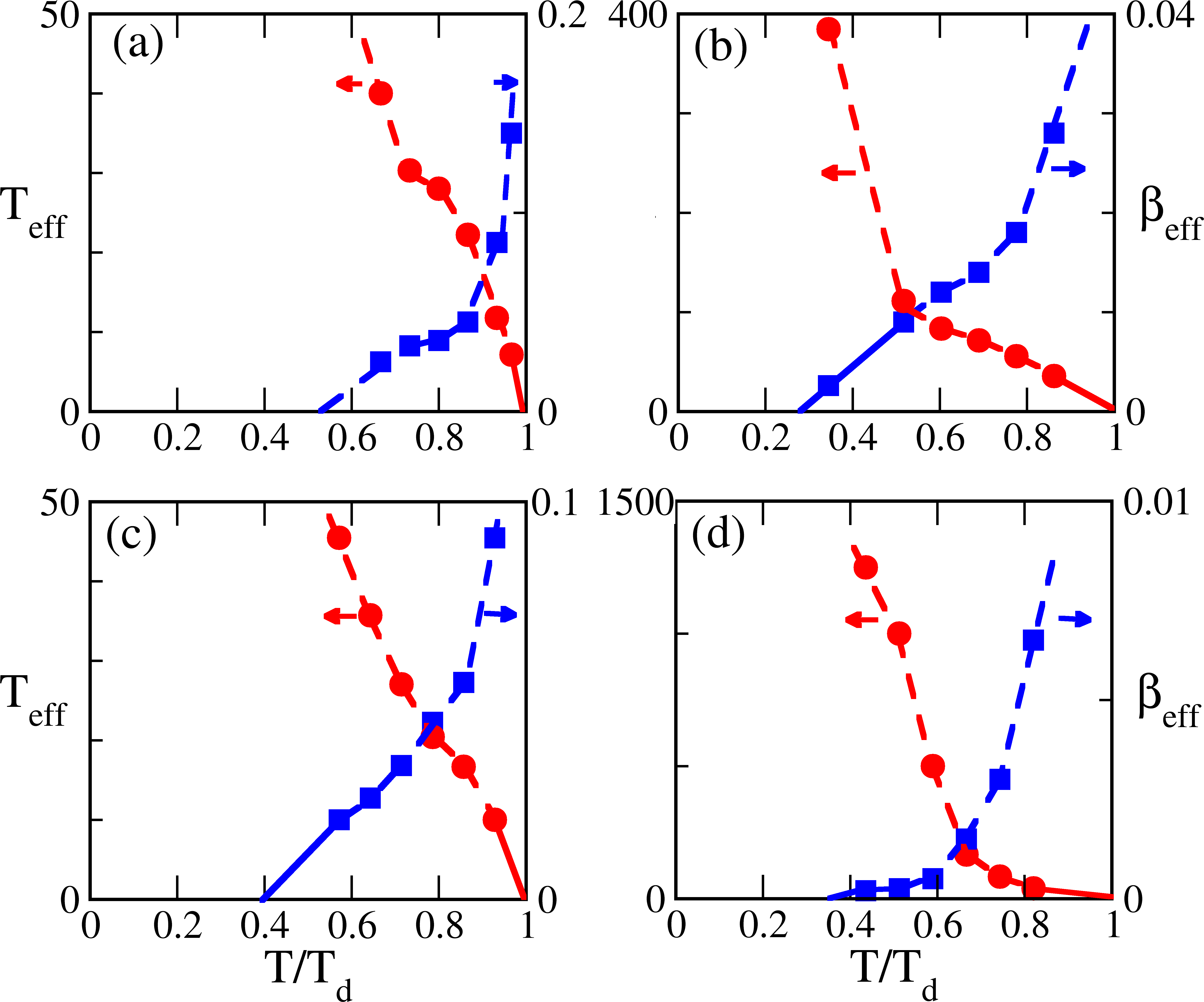}
\caption{{\it Effective temperature (and its  inverse) versus quench temperature:} Four-panel Figure for all transitions, showing   effective search temperature on the left vertical axis   versus $T$. For all transitions, the $T_{eff}$ vanishes linearly  $\sim T_d -T$  at a search-freezing  temperature $T_{d}$. The inverse   $\beta_{eff} (T)$ on the right vertical axis vanishes linearly  $\sim T-T_1$ at a search explosion temperature $T_1$. }    
\end{center}
\end{figure}

\begin{figure}[ht]
\begin{center}
\includegraphics[height=6.5cm, width=8.0cm]{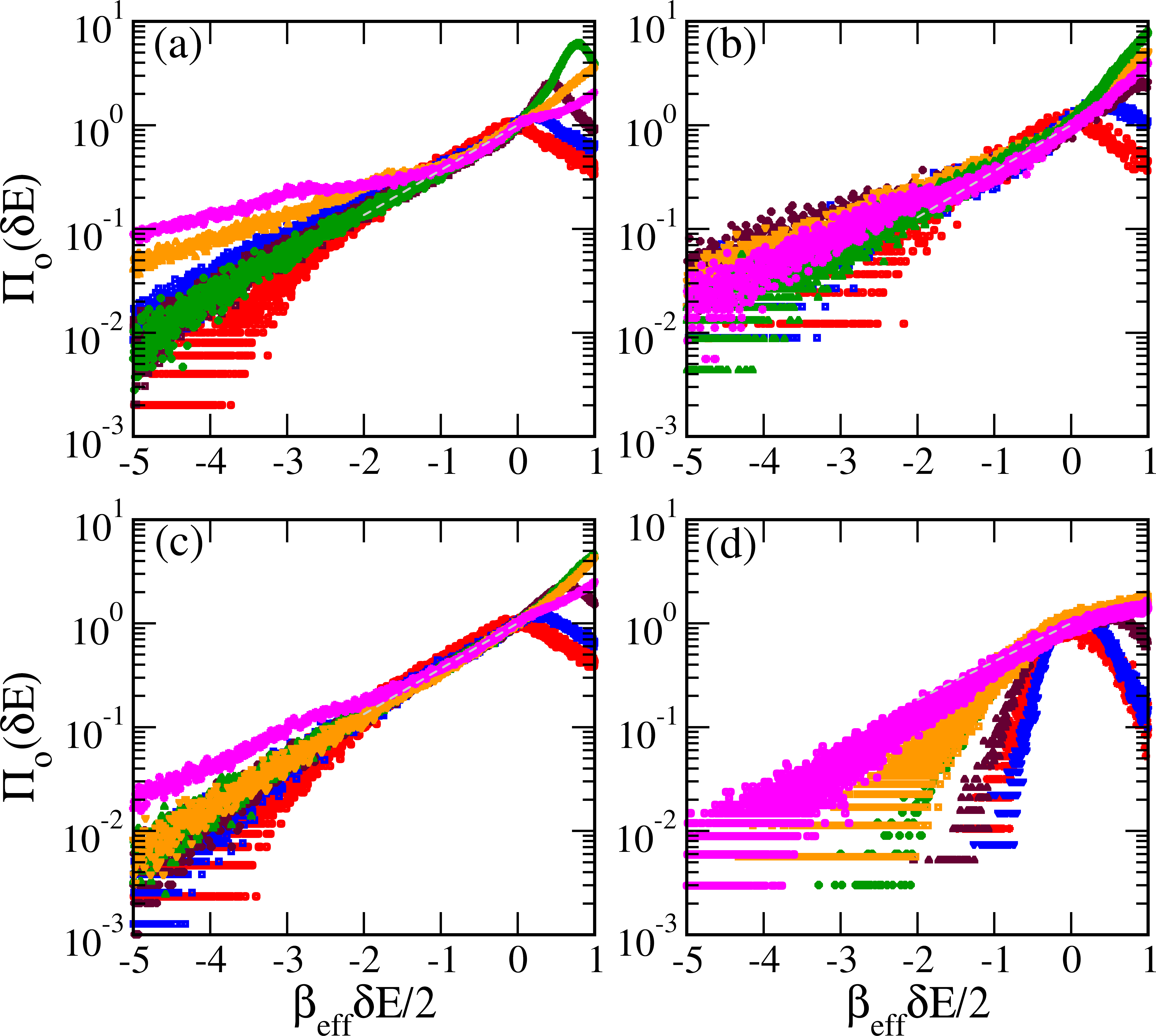}
\caption{{\it Four panel plot of scaled probability of energy change  $\Pi_0 $ versus $z= \beta_{eff} \delta E/2$.} The four transitions in  panels a),b),c),d) have respective slope mean values and standard deviations  of $1.000 \pm 0.045; ~1.025 \pm 0.036;~1.009 \pm 0.08;~0.850 \pm 0.085$. }
\end{center}
\end{figure}

 Fig 16 shows a central result, namely the search temperature  $T_{eff} (T)$ and its inverse, $\beta_{eff} (T)$ versus $T$. The left-axis search temperature $T_{eff} (T)$ intersects the temperature axis at an  extrapolated $T_{eff} (T_d) =0$,
defining a quench  temperature $T_d$.  The vanishing appears to be  linear,  $T_{eff} (T) \sim  T_d- T$.
 
 The mean conversion rate involves an integral over the heat releases of the distribution \cite{R39}. The mean time is then  a  singular exponential ${\bar t}_m \simeq t_0 e^{M / 4 T_{eff}} \simeq  t_0 e^{B_0 T_d/ (T_d -T) }$.  
 The Vogel Fulcher-Tammann form \cite{R5,R7,R8} thus emerges naturally   from a search temperature freezing inducing a rapid  arrest of PES cooling, and an entropy-barrier divergence. 
 
 Similarly, the inverse effective temperature $\beta_{eff} (T)$ on the right-axis of Fig 16 seems to go to zero  linearly $\sim (T-T_1)$  at a search explosion  $T_1$ where entropy barriers vanish.

Once again, the cubic-trigonal last panel is unusual, with $T_{eff}$ showing a smaller slope near $T_d$. 
If the linear slope  actually vanishes as  $T_{eff}(T) \sim (T- T_d)^2$  then that would yield `super-VFT' behaviour \cite{R5}, $t_m \sim e ^ {1/(T-T_d)^2}$.
 
Fig 17 shows the scaled ratio  of occurrence probability to its value at the origin,  $\Pi_0 \equiv P_0(\delta E,T)/ P_0 (0,T)$  versus $z \equiv \beta_{eff} \delta E /2 \simeq \delta S/2$.
The dashed white  lines  have slopes close to the predicted universal slope of  unity, as given in the Figure caption.

 The normalized probability $P_{MC}$ of a Monte Carlo spin-flip is the product of the occurrence probability  $P_0$ and  an MC acceptance factor with step functions,  
 
  \begin{equation}
 P_{MC} (\delta E , T) = \frac{P_0(\delta E, T)}{N_{MC}(T)} [ \theta (-\delta E) +  e^{-\delta E / T} \theta (\delta E)],
  \end{equation}
with $N_{MC} (T)$ a normalization constant. The  ratio of the MC probability and its value at the origin defines
$ \Pi_{MC} (\delta E, T)  \equiv [ P_{MC}  (\delta E, T)/ P_{MC} (0, T)]$. With 
$z \equiv \beta_{eff} (T) \delta E /2 $, 
 
 \begin{equation}
 \Pi_{MC} \simeq [e^{z} \theta (-\delta E) + e^{- z [ (2 T_{eff} /T) - 1]} \theta (\delta E)] / N_{MC} (T).
 \end{equation}

  Fig 18 shows a  log-linear plot  of $\Pi_{MC}$ along the positive axis  $ z=\beta_{eff} \delta E/2 > 0$. The  closeness  of data to  theoretical lines with  slope  $[1- 2 T_{eff}/T)] < 0$  is further evidence for PES.

\begin{figure}[ht]
\begin{center}
\includegraphics[height=6.5cm, width=8.0cm]{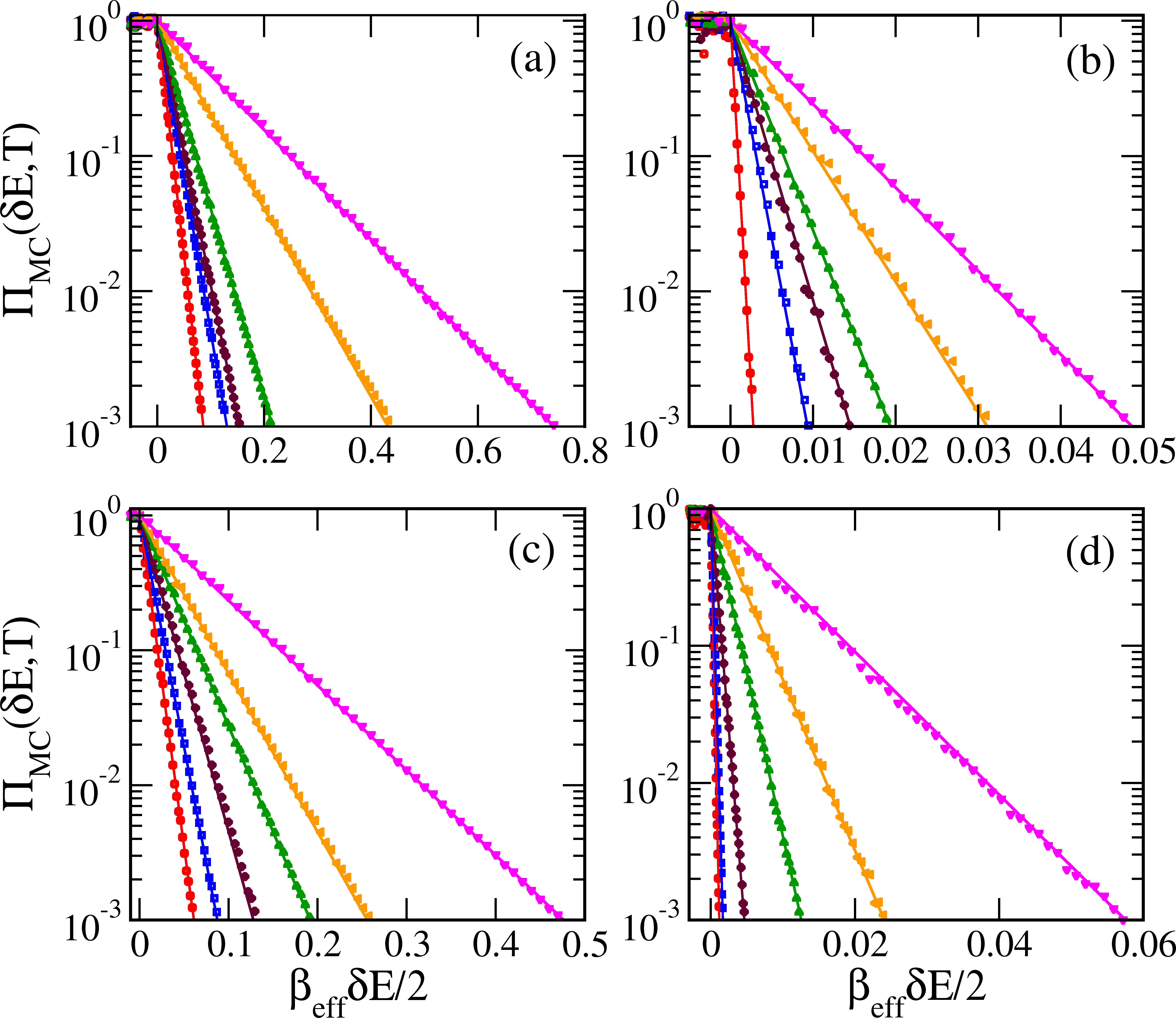}
\caption{{\it Four panel plot of  $\Pi_{MC}$ versus $z=\beta_{eff} \delta E/2 > 0$}.  The  light theoretical  line of  slope $[1- 2 T_{eff}/T)]$ has a reasonable match to data.}
\end{center}
\end{figure}

In conclusion all four transitions show PES signatures, lending support to the Partial Equilibration Scenario.

\section{PES  delays in simulations  and experiment}

  We  show that  delay data both  in simulations and in experiment,  are consistent with a picture of diverging entropy  barriers  from  linearly vanishing PES effective temperatures.

\subsection{VFT delays in 3D simulations}

\begin{figure}[ht]
\begin{center}
\includegraphics[height=6.5cm, width=8.0cm]{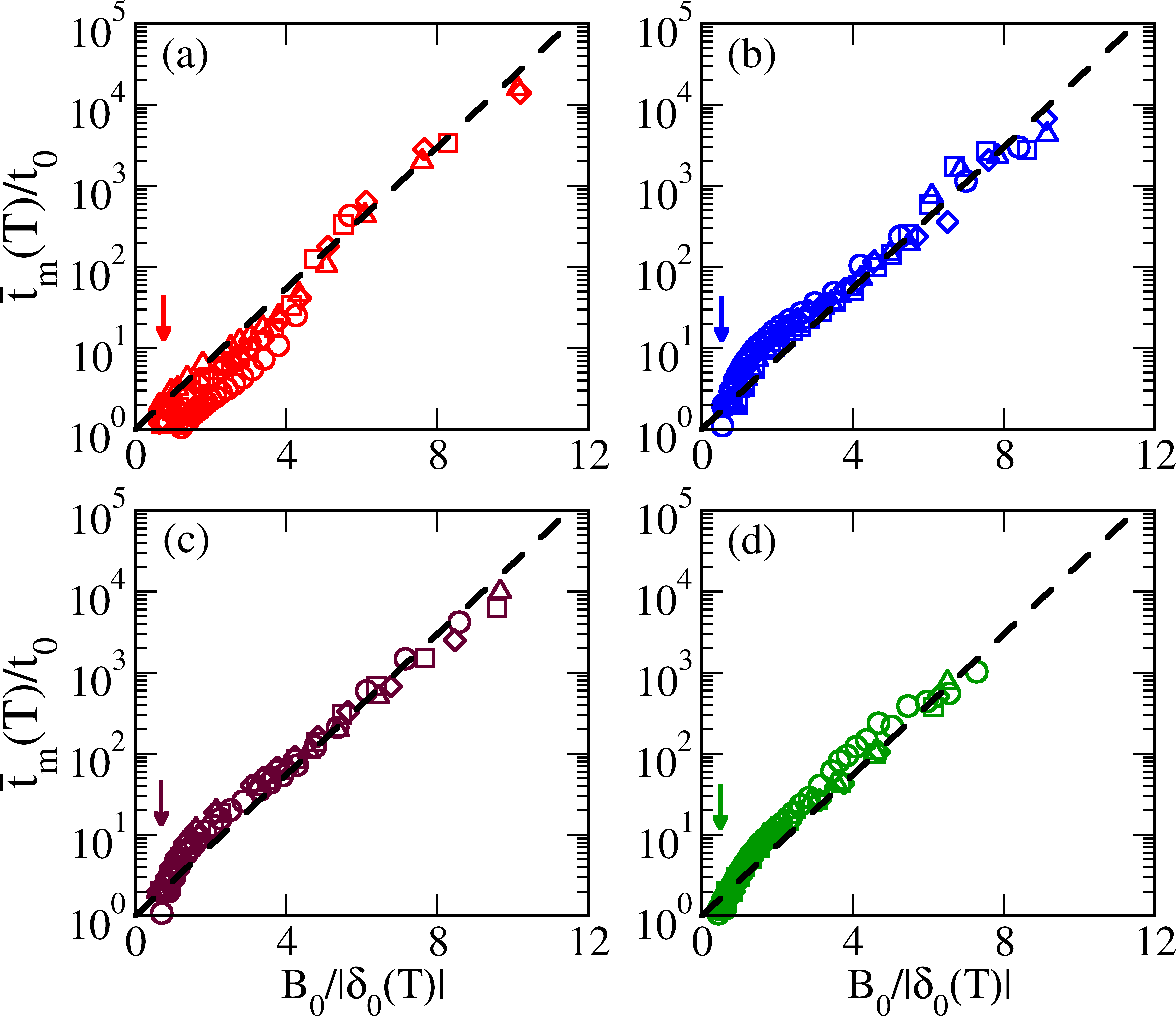}
\caption{{\it Vogel-Fulcher-like behaviour  in simulations:}
Four-panel plot of scaled data in simulations of log-linear  scaled ${\bar t}_m / t_0$ versus scaled    $B_0/ |\delta_0 (T)|$, for the four transitions  a),b),c), d), with each panel showing $E_0 =3,4,5,6$. Data clustering for small $|\delta_0 (T)|$, on the dashed line over three orders of magnitude is evidence for Vogel-Fulcher behaviour. For lower temperatures, there is a  peel-off  towards the  x-axis near  $T=T_1$ (downward arrow). } 
\end{center}
\end{figure}
 
 Vogel-Fulcher-Tamman temperature dependences can be  written  near $T_d$ as 
\begin{equation}
 {\bar t}_m (T) = t_0 e^{B_0 T_d/ |T- T_d|} = t_0 e^{B_0/|\delta_0 (T)|}
 \end{equation}
 where $t_0, B_0 T_d$ are the time and energy scales, for DW shifts of a lattice spacing.

  The logarithms of  VFT times can be written in two useful forms,  to extract constants $B_0,t_0$  from simulations and experiments. Thus   
 
  \begin{equation}
\frac{1}{\ln{ {\bar t}_m (T)}} =\frac{(|\delta_0 (T) |/B_0)}{ [ 1 +  (\log{t_0}) (|\delta_0 (T)| /B_0)]}
 \end{equation}
 and  

\begin{equation}
\ln{ {\bar t}_m (T)} = \log t_0 + [B_0/ |\delta_0 (T_0)| ].
 \end{equation}
 
For simulations, $B_0$ and $t_0$ can be extracted from data using  Equ (28) and Equ (29). Fig 19  shows  for all four transitions, the data in a  scaled  form of

\begin{equation}
\ln ({\bar t}_m / t_0) = B_0/|\delta_0 (T)|.
\end{equation}

There is  data clustering around  the Vogel-Fulcher straight line showing universality  over 3 orders of magnitude near $T_d$. There is also  a peel-off toward shorter times, near $T_1$ (downward arrow).

\subsection{VFT delays in martensitic alloys}

Delayed  athermal martensitic transformations in metallic alloys  have been tracked by  conversion diagnostics suited to the scale of the waiting times\cite{R16,R17,R18} such as  electrical resistivity drops; or surface optical or X-ray reflectivity. In pioneering experiments, Kakeshita et al \cite{R17}  used resistivity drops to detect the austenite to martensite delayed conversions for alloys  $Fe_x Ni_{1-x}$. The alloying percentage $100 ~x$ is $29.9, 31.6,  32.1 \%$, with start temperatures of $M_s = 239, 177, 148$ K.The delays discovered  were of macroscopically long times. Other work such as the Klemradt group\cite{R16,R18}  on $NiAl$ alloy delays, also rose rapidly: for temperature increments above the $M_s$ values  of $0.1$ K, $0.6$ K and $0.7K$, the delay  times went  from \cite{R18}  several seconds, to $10^4$ seconds, to forever. 

The data analysis of simulations is used again for experiment, as now described in more detail. To extract divergence temperatures $T_d$  from simulation or experimental data, we plot $1/ \ln(t)$ versus $T$, and  extrapolate\cite{R24} straight line segments to the x-axis (not shown). For FeNi data\cite{R17}  this yields $T_d \simeq 247, 187,158$ K,  well above $M_s$, with large fractional  delay windows  $|\delta_0 (M_s)| = |T_d-M_s|/ T_d = 0.03, 0.05, 0.06$.  Similarly extrapolation of data for the NiAl alloy\cite{R18} with $M_s = 282.2$K  yields  $T_d =283 K$ but with a smaller  window  $| \delta_0 (M_s)| \simeq 3 \times 10^{-3}$.

Another group considered NiTi alloys \cite{R20}, and argued that if the largest delay is at $T=T_0$, then for a long enough annealing time, conversions  should be seen at any $T$ in  a wide window $M_s <T< T_0$. However no conversion was detected, for holding at  $T =275.9$ above  $M_s= 274.3 K$,  for $t_h = 21$  days. The absence of conversion was attributed to a sparseness of (atomic) catalyst fields in facilitation-type delay models \cite{R20}.  In our picture this absence  could also be due to $T$ being  outside  the narrower  conversion  window $ M_s < T < T_d <T_0$,

Fig 20 shows measurement data  of conversion times $t$, in seconds, versus temperature $T$ in degrees Kelvin. The left column shows a  linear-linear plot like Equ (28) of   $1/ \ln { t}$ versus $|\delta_0(T)|$ to extract slope $1/B_0$  from Kakeshita data \cite{R17} for three alloys of FeNi (top panel); and from   Klemradt data\cite{R18} for a  NiAl alloy (bottom panel).  The right column shows a  linear-linear plot Equ (29)  of  $ \ln { t}$ versus $1/ |\delta_0 (T )$, using lines of the extracted slope $B_0$, to determine the  intercept $\ln t_0$ for FeNi (top panel)  and for  NiAl (bottom panel). Downward arrow marks $T_1=  $ for NiAl data.  The  `fragility' \cite{R8} parameter $B_0 T_d =1.23 K$, and a basic time scale for  DW hops is  $t_0 = 1$ second.

\begin{figure}[ht]
\begin{center}
\includegraphics[height=6.5cm, width=8.0cm]{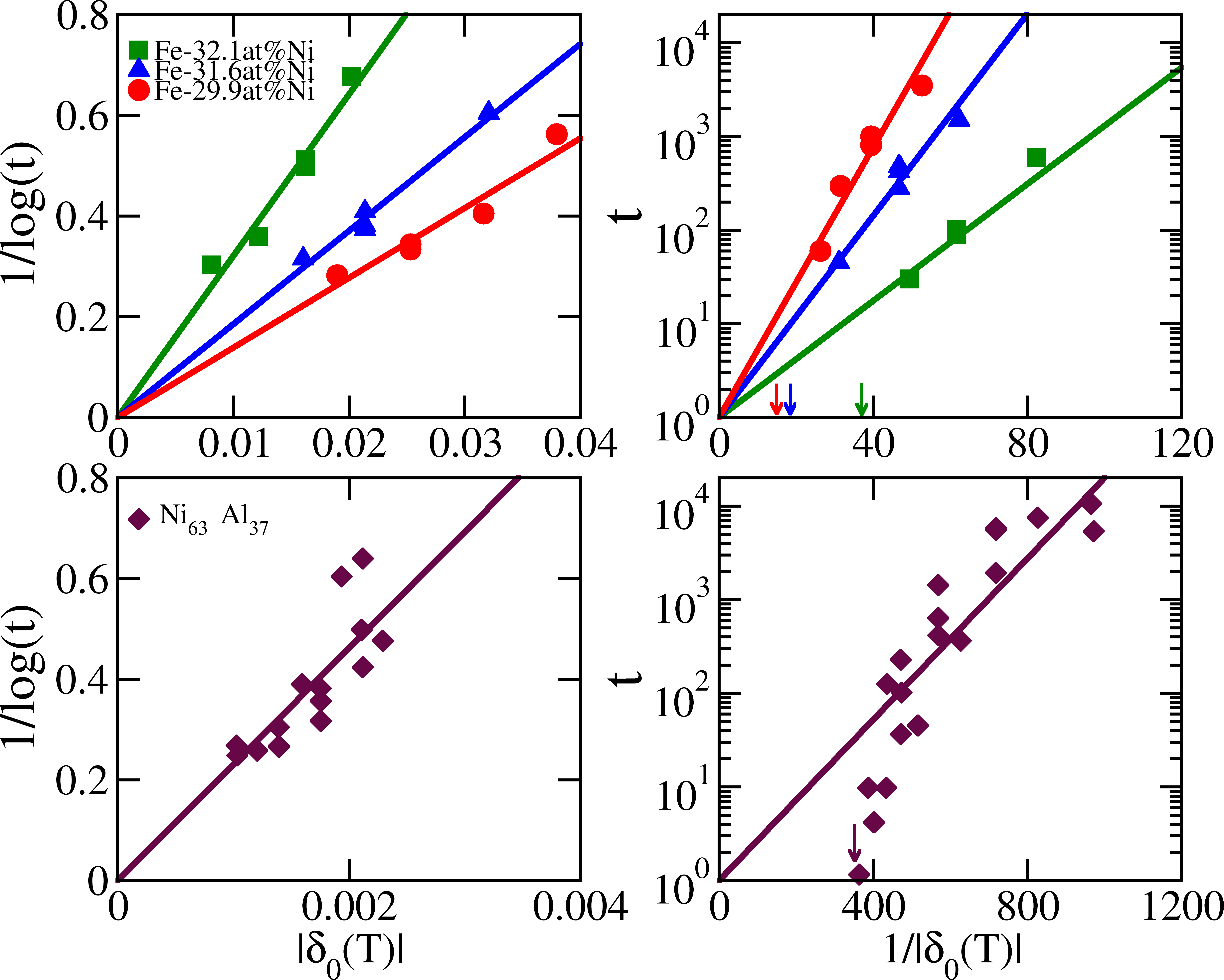}
\caption{{\it Vogel-Fulcher-like behaviour in experiment:} Left  column shows  data from both groups \cite{R16,R17,R18}, in a   $1/ \ ln(t)$ versus $ |\delta_0 (T)|$ plot. The slope $1/B_0$  is extracted  from a fit to $y = (1/B_0) x$.  Right  column again shows results of  both groups,  in a log-linear plot of  $\ln {\bar t}_m$ versus $1/|\delta_0 (T)$. The NiAl data show a downward deviation towards $T=M_s$  marked by the downward arrow.Using the extracted $B_0$ values, the intercepts  are found in a fit to $y = B_0 x + y_0$. }
\end{center}
\end{figure}

\begin{figure}[ht]
\begin{center}

\includegraphics[height=6.5cm, width=8.0cm]{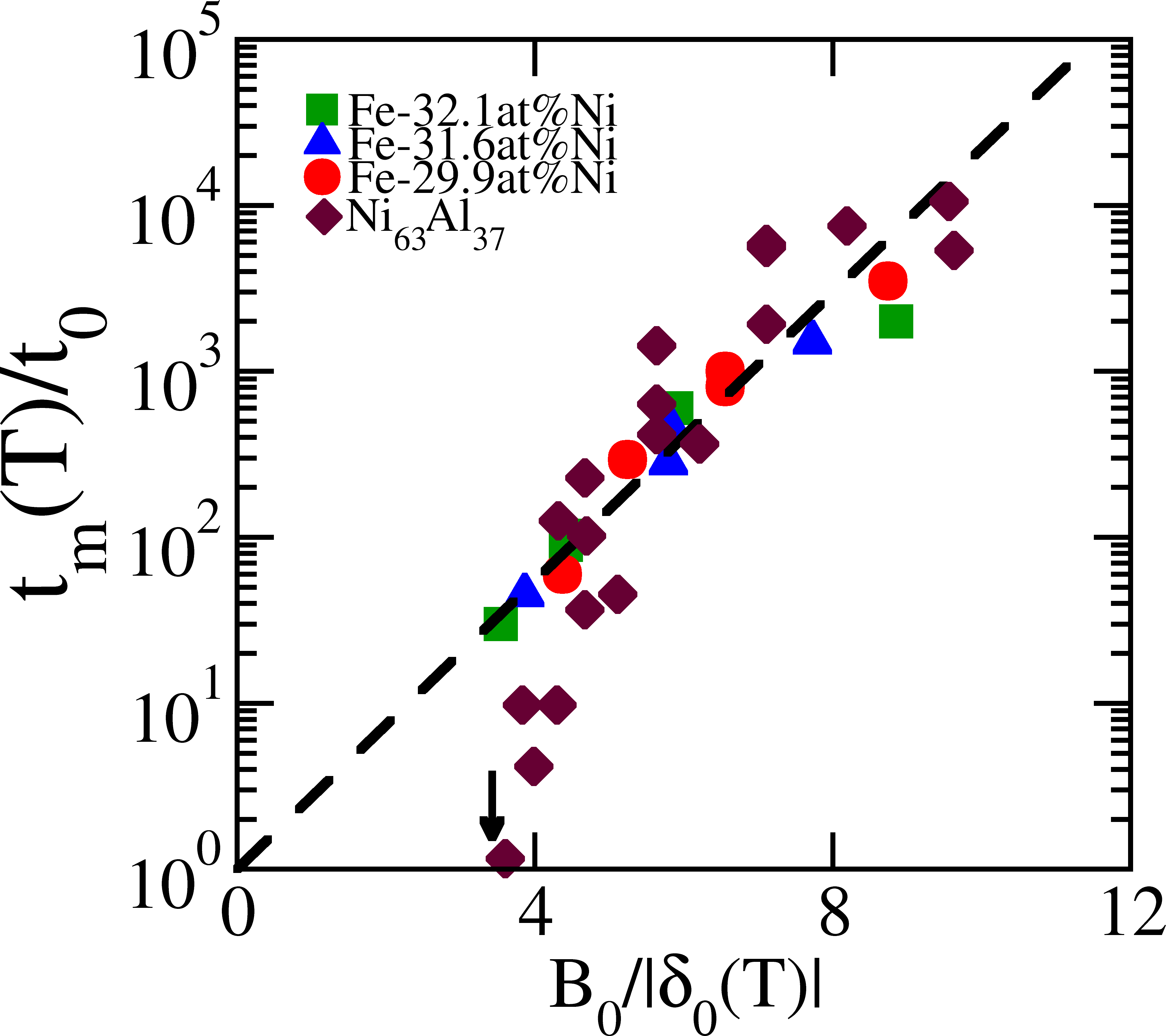}
\caption{{\it Entropy barrier collapse and divergence in scaled variables, for experiment:} Combined  data from experiments  in   linear-linear plot of $\log (t/t_0)$ versus $B/| \delta_0 (T)|$.  The dashed  straight line is the universal Vogel-Fulcher-Tammann form of the diverging entropy barrier.  Some  data show a downward turn for entropy barrier collapse. }
\end{center}
\end{figure}

Fig 21 shows that combined  experimental data \cite{R16,R17,R18}   cluster  around  the Vogel-Fulcher straight line of Equ (30), with  universality  over 3 orders of magnitude near $T_d$. For NiAl data\cite{R18},  there is a linear falloff on approaching $T_1$ (downward arrow), consistent with Fig 16. 

It would be interesting to get more data for these and other martensitic  alloys, through systematic quenches in steps of $1/|\delta_0(T)|$, over the entire delay range $M_s < T < T_d$ between barrier divergence and collapse. It would also be interesting to include  a $T_1$-like onset of  sluggishness,  in  fitting analyses of   glass-former viscosity data \cite{R7} 

\section{Summary and further work}

In this paper, we present Monte Carlo (MC) simulations,  on discrete-strain  Hamiltonians for four  3D structural transitions, under systematic temperature quenches from seeded austenite, to study austenite-to-martensite conversion times. The results and scenario are as follows.

 For athermal martensites, there are explosive conversions below a martensite start temperature $M_s$ so there are no barriers. Above this start temperature, entropy barriers emerge , and incubation time delays rise sharply towards a divergence temperature $T_d$. The entropy barrier collapse/ divergence,  is understood through  temperature-controlled phase-space bottlenecks. 

Partial equilibration  ideas provide an understanding of fast/ slow  times, based on effective temperatures  for  energy-lowering  searches. The inverse search temperature  vanishes linearly at $M_s$ or $1/ T_{eff} (T)  \sim |T- M_s| \rightarrow 0$. The search  temperature vanishes  linearly at $T_d$, or $T_{eff} (T) \sim |T-T_d| \rightarrow 0$. This  rapid search arrest explains the singular Vogel-Fulcher-Tammann form, extracted from martensitic experimental data.

Further simulations of crystallisation models \cite{R6}  could try to  record heat releases.  Further experiments could  record strain signals and intermittency\cite{R40} over the delay region $T_d > T> T_1$; and  over the tweed precursor \cite{R14,R25,R27} region above it  $T_0 > T> T_d$.  Non-stationary distributions of  energy changes in martensites could be measured through  concurrent acoustic, photonic or strain probes\cite{R40}. The  VFT temperature regime in glasses shows non-Debye frequency responses \cite{R9}: this might be more general. Finally,  quenches of complex oxides \cite{R41} near their  structural / functional  transitions, might yield  interesting PES signatures in functional variables, induced  by their coupling to  ageing strain domains. \\

Acknowledgements:\\
 It is a pleasure to thank Turab Lookman for helpful early conversations; and  Smarajit Karmakar  for valuable discussions on the glass transition.\\
 
{\bf Appendix A: PES signatures in the $m^4$ model:}\\

As a toy model illustration of the Section III procedure we consider  a 2D magnetization free energy  with double-well Landau term and  a Ginzburg term,   $F = E_0 \sum_{\vec r} [ f_L + f_G]$, Here $m$ is the single component OP, of  both signs, so $N_{OP} =1, N_V = 2$.  The Landau term  is
\begin{equation}
 f_L (m(\vec r))~ = \epsilon(T) m(\vec r)^2 + {m(\vec r) ^4 }/ 2
\end{equation}
where $\epsilon(T) \equiv (T -T_c)/T_c$, with all energies/ temperatures  scaled  in the physical  transition temperature, so the scaled $T_c =1$. The Ginzburg term is
$f_G = \xi_0 ^2 [{{\Delta  m}(\vec r)}^2]$.

\begin{figure}[ht]
\begin{center}
\includegraphics[height=3.6cm, width=8.5cm]{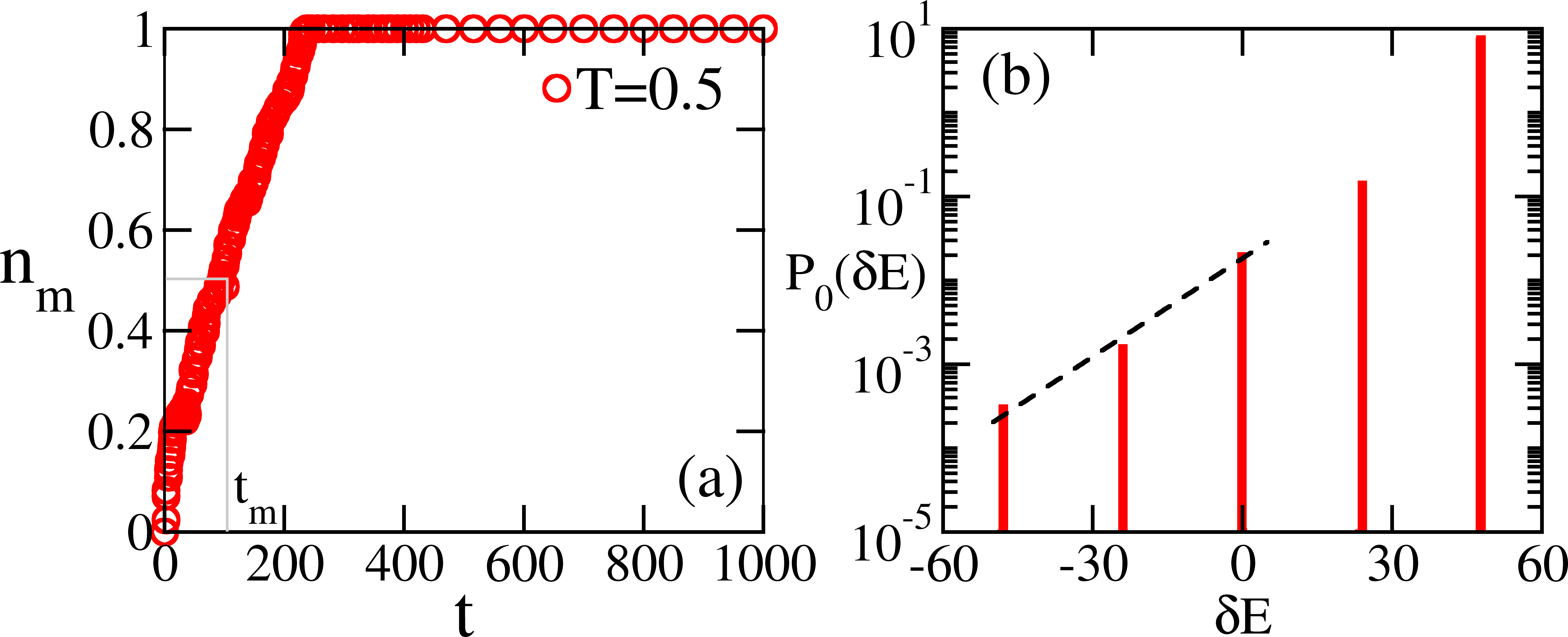}
\caption{{\it  Magnetization fraction and PES distribution:}  ~ a) Linear-linear plot of magnetization fraction  $n_m (t)$ versus time after a quench to  $T=0.5$. The  curve at $n_m (t_m) =1/2$ defines a halfway time $t_m(T)$, that here is $t_m = 107$ MC steps.  ~  b) Log-linear histogram  of normalized probability of heat release  $P_0 (\delta E, T)$ versus $\delta E$ after the quench, recording  energy changes up to  waiting  times $t _w= t_m (T)$.
 Here the slope is  $\beta_{eff}/2  = 0.09$, or $T_{eff} =5.6$. }. 
\end{center}
\end{figure}

Domain-walls are  solitonic solutions with a $\tanh$ profile,  interpolating between  flat OP variants of opposite sign, within a   DW  thickness  $ \xi_0 \sim 1$. 
The OP can be written  as a magnitude $|m|$ times a variant `spin'  $S(\vec r)=\pm 1$ of unit length. 
The nearly flat magnitude in the domains is approximated by the mean field OP,
\begin{equation}
m(\vec r) \equiv |m(\vec r) | S(\vec r)  \rightarrow {\bar m} (T) S(\vec r) 
\end{equation}
where ${\bar m}(T) = |\epsilon (T)|^{1/2}$. Compare Equ (7).

The mean-field Landau free energy is
\begin{equation}
f_L(T)  = {\bar m} (T)^2 g_L(T) ;~ g_L =  -\frac{1}{2} |\epsilon(T)| <0.
\end{equation}
where $g_L (T_c)=0 $. Compare Equ (10).

Substituting Equ (32) in Equ (31)  yields  the DW coordinate space Hamiltonian, with $D_0 \equiv 2 {\bar m} ^2 E_0/T$,   
\begin{equation}
\beta H = (D_0/2) \sum_{\vec r} [-| g_L(T)| S(\vec r)^2 +\xi_0 ^2 (\Delta S(\vec r))^2 ],
\end{equation}
although in this case,  $S(\vec r)^2 =1$ at all sites. With $S(\vec r) = \sum_{\vec k} e^{i\vec k . \vec r} S(\vec k) / \sqrt{N}$, the Hamiltonian in Fourier space  is
\begin{equation}
\beta H = (D_0 /2) \sum_{\vec k } [-| g_L (T)| + \xi_0 ^2 {\vec K} (\vec k) ^2] |S(\vec k)|^2.
\end{equation}
Compare Equ (13) and Equ (14).

We can do  MC simulations with this T-dependent,  Ising-variant effective Hamiltonian. The parameters are $N = 64^2, \xi_0 ^2 =1, N_{run} =10$, with holding times $t_h = 10^3$. 
The  spin-flips are  $S= \pm 1 \rightarrow \mp 1$. Fig 22a  shows the magnetization fraction $n_m (t)$ versus waiting time $t_w$ analogous to Fig 3, but here with a gradual rise, and no  incubation behaviour.

We record $\{ \delta E\}$ for every spin flip up to an OP marker event time $t_w =t_m$ when $n_m (t_m) =1/2$. For nearest-neighbour couplings on a square lattice, the energy changes  $\delta E$  will be discrete. Fig 22b shows  the  log-linear $P_0 (\delta E, t_w =t_m)$ versus discrete energy changes $\delta E$.  The spike heights  decrease linearly with energy changes,  consistent with PES.  Compare Fig 15.

\begin{figure}[ht]
\begin{center}
\includegraphics[height=3.8cm, width=8.5cm]{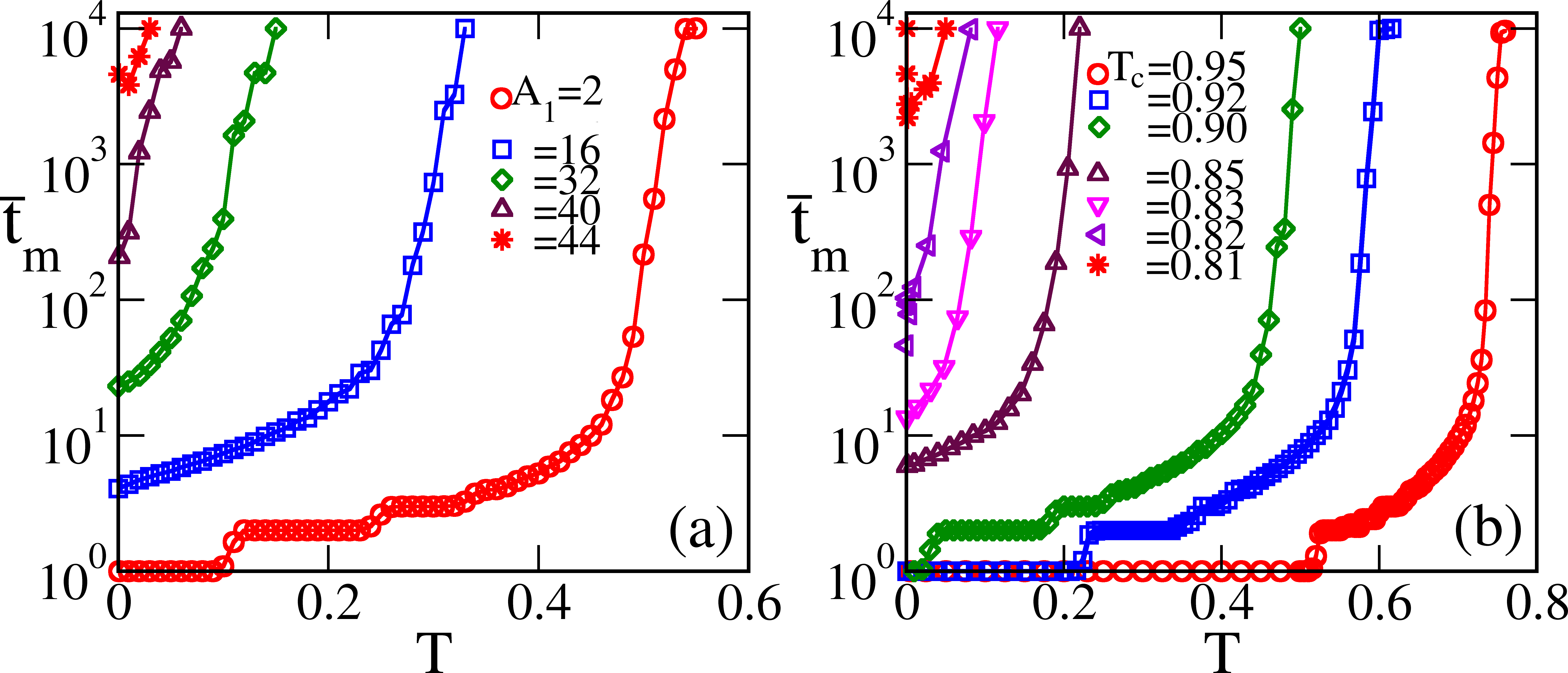}
\caption{{\it  Tetragonal-orthorhombic  TTT curve, showing both explosive and incubated conversions:} Figures show TTT curves  as log-linear plots of time versus quench temperature. Flat regions along the x-axis denote single time-step conversions up to end points $T=T_1$, defining athermal regime materials. The $T_1$  depend on the elastic constants $A_1$ and spinodal temperatures $T_c$. For some parameters, $T_1$ is driven to zero, and the TTT curves intersect the y-axis. These  materials are  in a `mixed' regime. See text.  }
\end{center} 
\end{figure}

{\bf Appendix B:  Athermal phase diagram:}\\

\begin{figure}[ht]
\begin{center}
\includegraphics[height=7.0cm, width=8.0cm]{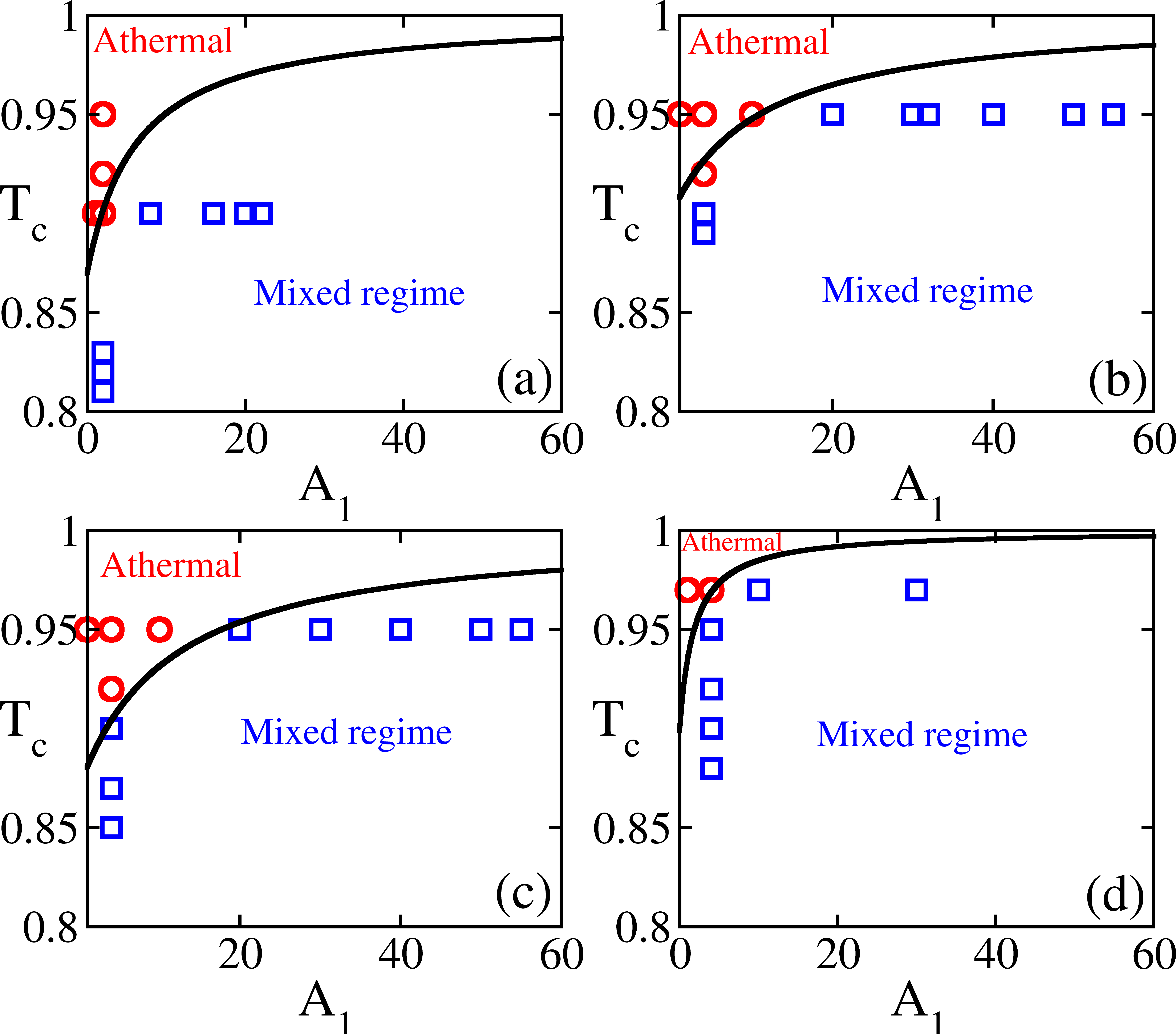}
\caption{{\it  Phase diagrams for athermal/ mixed behaviour:} Four-panel Figure showing  linear-linear plots of  $T_c$ versus $A_1$ solid line the theoretical boundary between athermal and mixed  materials. Simulation data for athermal (open circles) and mixed (open squares)   behaviour,  are  consistent with the  theoretical curve.   }
\end{center} 
\end{figure}

Fig 23 shows data for the martensitic  tetragonal-orthorhombic  transition, for several  $T_c, A_1$, showing both  athermal and  non-athermal or mixed behaviour \cite{R11,R23,R24,R25}. For the athermal regime, Temperature-Time-Transformation (TTT)  curves have  flat lines along the temperature axis where there are  immediate, explosive  conversions  for $T<T_1$. The embryo or droplet is small in coordinate space, and so is broad and flat in Fourier space. With $|{\vec S}(\vec k)|^2$ approximated by a constant,
the Hamiltonian energy of  Equ (18) is $E/D_0  \sim  \sum_{\vec k} \epsilon (\vec k)$  then involves averages of terms over the Brillouin Zone, denoted   by square brackets. The droplet energy  is
\begin{equation}
E (T) / D_0 \simeq - |g _L (T)|  + \{ \xi_0 ^2 [ K^2 (\vec k)]  + (A_1 /2 ) [U_{\ell, \ell} (\hat k)]\}.
\end{equation}

The energy vanishing $E(T_1)=0$ defines $T=T_1$. For small values in  a Taylor expansion, and with coefficient $b(T)$ as defined in the text,
\begin{equation}
 T_1 (A_1, T_c) \simeq   \{ + |g _L (0)| -  \xi_0 ^2 [ K^2 (\vec k)] - (A_1 /2 ) [ U_{\ell, \ell} (\hat k)]  \} / |b(0)|.
\end{equation}
 We define `athermal' behaviour as a nonzero $T_1$; explosive conversions for $T<T_1$; and  incubation delays in the window $T_1 < T < T_d < T_0$. The behaviour not precisely athermal, is called `mixed',  with conversions occurring gradually, without  flat incubations.
Vanishing of   the athermal case start temperature $T_1 (A_1,T_c) =0$ then determines a phase boundary.

The four-panel Fig 24  plot of $T_c $ versus $A_1$,  shows the phase boundary.  Above the phase boundary  the system is purely athermal, while  below the phase boundary where $T_1 =0 $, is a mixed regime.  Data  from Fig 23 and other TTT  diagrams are seen to be consistent with the theoretical phase boundary. Fig 23 shows that for some parameters, curves on the upper left, have  a fall and then a rise with temperature, like a `U shape, or downward `nose' .  For the large $E_0 =3$ used, the shape is distorted,  but  for smaller $E_0<1$ the U shape is more well defined \cite{R26}. The  shape  is from an Arrhenius activation over a temperature-dependent  energy barrier\cite{R42}.


\begin{thebibliography}{2004}
\bibitem{R39}
The mean equilibration rate is  the average over an inter-shell rate $1/t(\delta E)$ over all heat releases. With a normalized PES probability, 
$1/{\bar t} ={ \int_{-\infty}^{0} d\delta E~ P_0 (\delta E, T)/ t(\delta E)}$. For slow intershell variation $1/t(\delta E) \simeq 1/t_0$, the mean rate is proportional to the acceptance fraction over energy decrements. Taking the  PES probability, as a shifted gaussian with peak mean  energy change $M >0$, and variance $\sigma^2 = 2M T_{eff}$, the  acceptance fraction is a complementary error function, whose asymptotic behaviour yields $t_0 /{\bar t} \simeq e^{-M/4T_{eff}}$.  Dropping constants, the martensitic mean conversion time near $T_d$ is written 
simply as ${\bar t}_m (T) = t_0 ~ e^{1/T_{eff}(T)}$.
\bibitem{R42}
Activated times $t = e^{U(T)/T}$ over rising energy barriers $U(T) = a + b T^2$ diverge at $T=0$. The times fall for $T$ increasing from zero, have a minimum, and  then rise exponentially  at large $T$. The  minimum in $U(T)/T$ is  at $T=T^{*} \equiv \sqrt{a/b}$. Close to $T^{*}$, the TTT curve   has an `isothermal' U-shape or nose:  $ t(T) /t(T^{*}) = e^{2(ab)^{1/2} [(T/T^{*})-1)^2]}$.
\end{thebibliography}
\end{document}